\documentclass[aps,prl,a4paper,10pt,twocolumn,showpacs,floatfix,superscriptaddress,amsmath,amsfonts,amssymb,preprintnumbers,longbibliography,nofootinbib]{revtex4-1}

\usepackage{float}

\usepackage{dcolumn,graphicx,color,booktabs,microtype,afterpage}
\graphicspath{{./}{figure/}}
\usepackage[charter,greekuppercase=italicized]{mathdesign}
\usepackage{sidecap}
\renewcommand{\tablename}{Table}
\makeatletter\renewcommand{\fnum@figure}[1]{\figurename~\thefigure.~}\makeatother
\makeatletter\renewcommand{\fnum@table}[1]{\tablename~\thetable.}\makeatother

\newcount\hh \newcount\mm
\hh=\time \divide\hh by 60
\mm=\hh \multiply\mm by 60 \mm=-\mm
\advance\mm by \time
\def\now{\number\hh:\ifnum\mm<10{}0\fi\number\mm}

\usepackage[colorlinks,plainpages=false,linkcolor=blue,urlcolor=blue,citecolor=blue,pdfpagemode=UseNone,pdfstartview=FitBH]{hyperref}

\begin{document}
	
	\makeatletter\renewcommand{\ps@plain}{%
		\def\@evenhead{\hfill\itshape\rightmark}%
		\def\@oddhead{\itshape\leftmark\hfill}%
		\renewcommand{\@evenfoot}{\hfill\small{--~\thepage~--}\hfill}%
		\renewcommand{\@oddfoot}{\hfill\small{--~\thepage~--}\hfill}%
	}\makeatother\pagestyle{plain}
	
\preprint{\textit{Preprint: \today, \now. }} 

\title{Anomalous Hall resistivity and possible topological Hall effect in the EuAl$_4$ antiferromagnet} 

\author{T.\ Shang}
\email{tian.shang@psi.ch}
\affiliation{Key Laboratory of Polar Materials and Devices (MOE), School of Physics and Electronic Science, East China Normal University, Shanghai 200241, China}
\affiliation{Laboratory for Multiscale Materials Experiments, Paul Scherrer Institut, CH-5232 Villigen PSI, Switzerland}

\author{Y.\ Xu}
\email{yangxu@physik.uzh.ch}
\affiliation{Physik-Institut, Universit\"{a}t Z\"{u}rich, Winterthurerstrasse 190, CH-8057 Z\"{u}rich, Switzerland}

\author{D.\ J.\ Gawryluk}
\affiliation{Laboratory for Multiscale Materials Experiments, Paul Scherrer Institut, CH-5232 Villigen PSI, Switzerland}

\author{J. Z. Ma}
\affiliation{Swiss Light Source, Paul Scherrer Institut, CH-5232 Villigen PSI, Switzerland}

\author{T.\ Shiroka}
\affiliation{Laboratorium f\"ur Festk\"orperphysik, ETH Z\"urich, CH-8093 Zurich, Switzerland}
\affiliation{Laboratory for Muon-Spin Spectroscopy, Paul Scherrer Institut, CH-5232 Villigen PSI, Switzerland}

\author{M.\ Shi}
\affiliation{Swiss Light Source, Paul Scherrer Institut, CH-5232 Villigen PSI, Switzerland}

\author{E.\ Pomjakushina}
\email{ekaterina.pomjakushina@psi.ch}
\affiliation{Laboratory for Multiscale Materials Experiments, Paul Scherrer Institut, CH-5232 Villigen PSI, Switzerland}

\date{\today}

\begin{abstract}
We report the observation of anomalous Hall resistivity in single crystals 
of EuAl$_4$, a centrosymmetric tetragonal compound, which
exhibits coexisting antiferromagnetic (AFM) and charge-density-wave (CDW) orders with onset at
$T_\mathrm{N} \sim 15.6$\,K and $T_\mathrm{CDW} \sim 140$\,K, respectively. 
In the AFM state, when the magnetic field is applied along the $c$-axis 
direction, EuAl$_4$ undergoes a series of metamagnetic transitions. 
Within this field range, 
we observe a clear hump-like anomaly in the Hall resistivity, 
representing part of the anomalous Hall resistivity. 
By considering different scenarios, we conclude that 
such a hump-like feature
is most likely a manifestation of the topological Hall effect, 
normally occurring in noncentrosymmetric materials known to host nontrivial topological spin textures. 
In view of this, EuAl$_4$ would represent a rare case where the 
topological Hall effect not only arises in a centrosymmetric 
structure, but it also coexists with CDW order. 
\end{abstract}


\maketitle

\emph{Introduction.} 
The Hall effect, involving either the charge- or the spin degree of 
freedom,
is at the research frontier due to its  
possible applications in spintronic devices~\cite{Nagaosa2013,Sinova2015,Hirohata2020}.
In the charge channel, the Hall resistivity $\rho_{xy}$ in a magnetic 
material can generally be decomposed into two components, 
$\rho_{xy} = \rho_{xy}^O + \rho_{xy}^A$, 
where $\rho_{xy}^O$ and $\rho_{xy}^A$ represent the ordinary- and the 
anomalous Hall resistivity, respectively. Further on, $\rho_{xy}^A$ can 
be split into a conventional anomalous Hall term $\rho_{xy}^{A'}$, 
mostly determined by the magnetization $M$ and the electrical resistivity 
$\rho_{xx}$, and a topological Hall term $\rho_{xy}^T$. 
The topological Hall effect is considered  
the hallmark of spin textures with a finite scalar spin chirality in 
real space~\cite{neubauer_topological_2009,gayles_dzyaloshinskii-moriya_2015,kanazawa_large_2011,franz_real-space_2014,schulz_emergent_2012,qin_emergence_2019,matsuno_interface-driven_2016,kurumaji_skyrmion_2019,kanazawa_critical_2016,fujishiro_topological_2019,gobel_topological_2020,vistoli_giant_2019}.
Such topological spin textures exhibit a nonzero Berry phase, which 
acts as an effective magnetic field  
and gives rise to topological Hall resistivity, namely $\rho_{xy}^T$. 
Among the notable examples in this regard are the noncentrosymmetric MnSi 
and analog compounds~\cite{neubauer_topological_2009,gayles_dzyaloshinskii-moriya_2015,kanazawa_large_2011,franz_real-space_2014}, 
where $\rho_{xy}^T$ is caused by magnetic skyrmions.

The tetragonal BaAl$_4$-type structure represents the prototype
for many binary- and ternary derivative compounds~\cite{Kneidinger2014}.
The research on tetragonal AE(Al,Ga)$_4$ (AE = Sr, Ba, and Eu) materials 
was recently reinvigorated by the discovery of nontrivial band topology 
in BaAl$_4$, where also a giant magnetoresistance (MR) was observed~\cite{wang_crystalline_2020}. Both BaAl$_4$ and BaGa$_4$ exhibit metallic behavior without showing any phase transition, while SrAl$_4$ shows a charge-density-wave (CDW) and a structural phase transition at $T_\mathrm{CDW} \sim 250$\,K and $T_\mathrm{S} \sim 90$\,K, respectively~\cite{Nakamura2016}. 
Unlike its nonmagnetic counterparts, EuGa$_4$ is an antiferromagnet below 
$T_\mathrm{N} \sim 16.5$\,K, while EuAl$_4$ undergoes a series of 
antiferromagnetic (AFM) transitions in its CDW ordered state~\cite{nakamura_magnetic_2013,araki_charge_2013,nakamura_unique_2014,nakamura_transport_2015,shimomura_lattice_2019,Kobata2016}. 
Clearly, in the Eu(Al,Ga)$_4$ family, the $4f$ electrons bring new 
intriguing aspects to the topology. 
	
%
	\begin{figure}
		\includegraphics[width=0.45\textwidth]{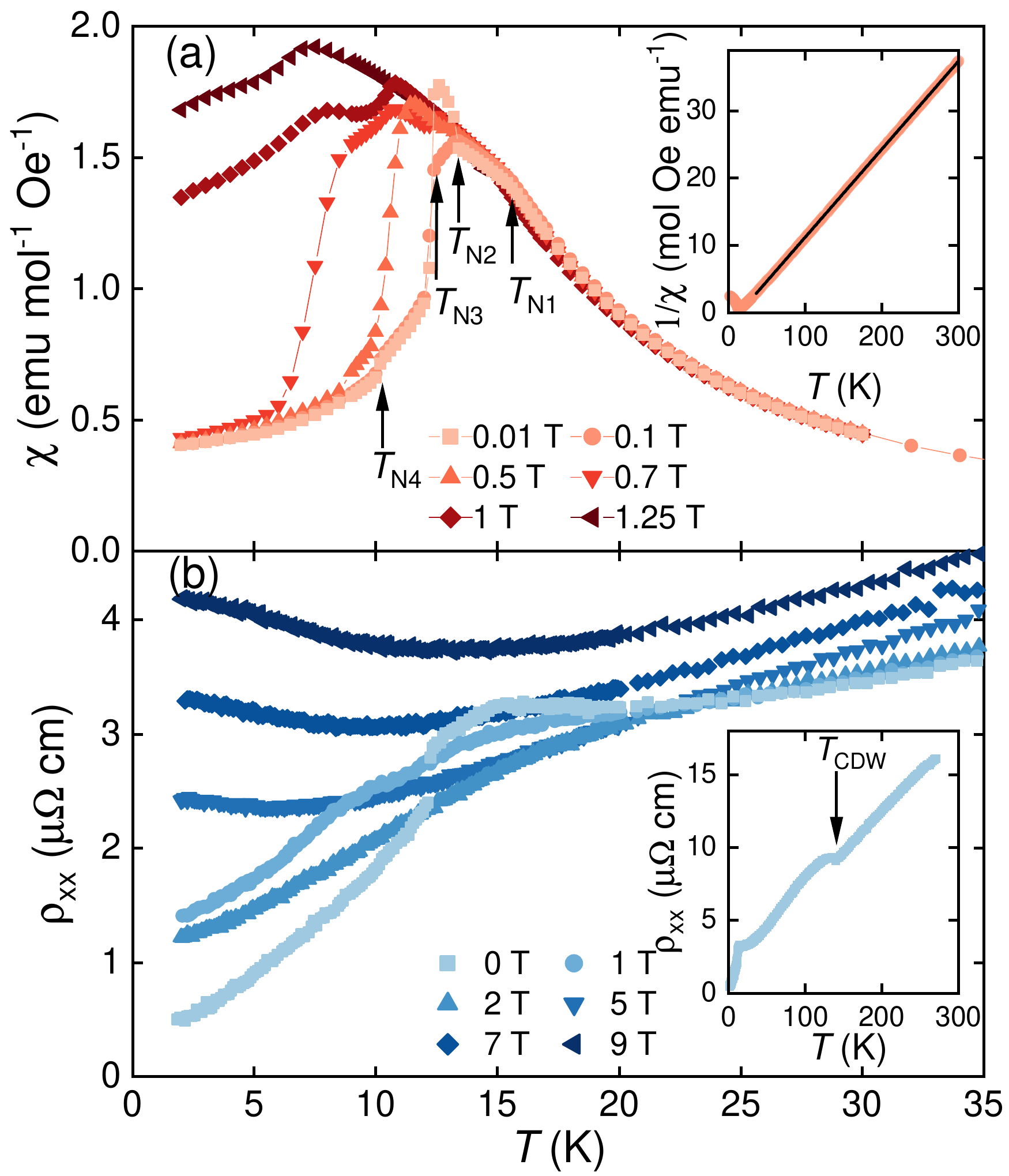}
		\centering
		\vspace{-2ex}%
		\caption{\label{fig:Tdepen}Temperature dependence of (a) magnetic 
		susceptibility $\chi(T,H)$ and (b) electrical resistivity 
		$\rho_{xx}(T,H)$ of EuAl$_4$, measured in various applied magnetic fields. 
		The arrows in (a) denote the four different magnetic transitions. 
		The insets show the 0.1-T inverse susceptibility $\chi(T)^{-1}$
		(top) and the zero-field $\rho_{xx}(T)$ up to $\sim 300$\,K
		(bottom). 
		The solid line in the top inset is a fit to the Curie-Weiss
		law, while the arrow in the bottom inset marks the CDW transition 
		occurring at $T_\mathrm{CDW} \sim 140$\,K.}
	\end{figure}
%
	
Most of the previous work on Eu(Al,Ga)$_4$ has focused on their temperature-dependent 
properties,
with the electrical transport properties under applied magnetic fields being 
somewhat overlooked~\cite{nakamura_magnetic_2013,araki_charge_2013,nakamura_unique_2014,nakamura_transport_2015,shimomura_lattice_2019,Kobata2016}. 
Here, we report the observation of a hump-like anomaly in the Hall 
resistivity of EuAl$_4$ single crystal. Since such anomaly appears  
in the magnetic field region where a series of metamagnetic transitions take place, most likely it is caused by the topological spin textures. 
Yet, we consider also the possibility of a regular origin of such anomaly 
in the Hall resistivity.

\emph{Experimental details.} Single crystals of EuAl$_4$ were grown by a molten Al flux method.
The crystals were checked by powder x-ray diffraction (XRD) measured 
using a Bruker D8 diffractometer. No extraneous phases could be 
identified in the XRD pattern, while Rietveld refinement confirmed 
the tetragonal crystal structure ($I4/mmm$, No.\ 139) with 
lattice parameters $a = b = 4.400$\,\AA{} and $c = 11.167$\,\AA{}. 
Magnetization and electrical resistivity measurements were performed in 
a Quantum Design MPMS and PPMS system, respectively. For the resistivity 
measurements, the electric current was applied in the $ab$-plane, while 
the magnetic field was applied along the $c$-axis.
To avoid spurious resistivity contributions due to misaligned Hall probes, all the resistivity measurements were performed in both positive and negative magnetic fields. Then, in the case of the Hall resistivity $\rho_\mathrm{xy}$, the spurious longitudinal contribution was removed by an anti-symmetrization procedure, i.e., $\rho_\mathrm{xy}(H) = [\rho_\mathrm{xy}(H) - \rho_\mathrm{xy}(-H)]/2$. Whereas in the case of the longitudinal electrical resistivity $\rho_\mathrm{xx}$, the spurious transverse contribution was removed by a symmetrization procedure, i.e., $\rho_\mathrm{xx}(H) = [\rho_\mathrm{xx}(H) + \rho_\mathrm{xx}(-H)]/2$.

\begin{figure}
	\includegraphics[width=0.45\textwidth]{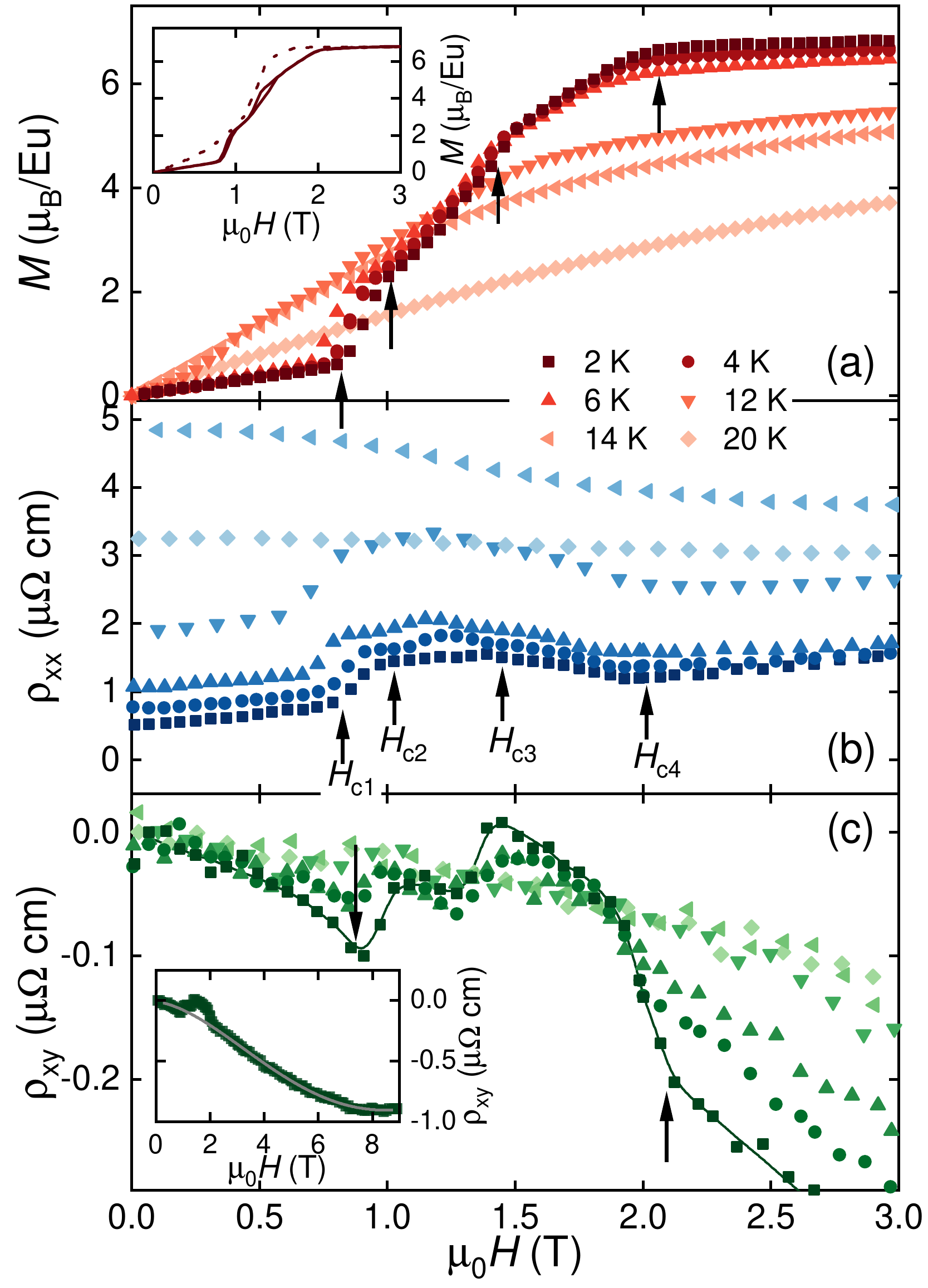}
	\centering
	\vspace{-2ex}%
	\caption{\label{fig:Hdepen}Field dependence of (a) magnetization 
	$M(H,T)$, (b) electrical resistivity $\rho_{xx}(H,T)$, and (c) 
	Hall resistivity $\rho_{xy}(H,T)$ of EuAl$_4$, collected at 
	various temperatures. The arrows in panels (a) and (b) mark the 
	saturation field ($H_\mathrm{c4}$) and the three critical fields 
	($H_\mathrm{c1}$, $H_\mathrm{c2}$, $H_\mathrm{c3}$) where EuAl$_4$ 
	undergoes metamagnetic transitions. The arrows in (c) denote the 
	upper- and lower field limit, where the hump-like anomaly appears 
	in the $\rho_{xy}(H,T)$ data. The inset in (a) shows the 2-K 
	magnetization, 	with the magnetic field applied along the $c$-axis 
	(solid line) or in the $ab$-plane (dashed line). 
	The inset in (c) shows the 2-K $\rho_{xy}(H)$ resistivity data up to 9\,T, 
	with the solid line being a polynomial fit.}
\end{figure}

\emph{Results and dicussion.} The temperature dependence of the magnetic 
susceptibility $\chi(T,H)$ and electrical resistivity $\rho_{xx}(T,H)$ 
of EuAl$_4$, measured under various magnetic fields, are shown in 
Fig.~\ref{fig:Tdepen}. Four successive antiferromagnetic transitions can 
be clearly identified in the $\chi(T)$ measured in a small magnetic field 
($< 0.1$\,T), as indicated by the arrows in Fig.~\ref{fig:Tdepen}(a). 
The zero-field-cooling- and field-cooling magnetic susceptibilities are 
practically identical, thus confirming the AFM nature of these transitions.  
The transition temperatures, $T_\mathrm{N1} \sim 15.6$, 
$T_\mathrm{N2} \sim 13.4$, $T_\mathrm{N3} \sim 12.6$, and 
$T_\mathrm{N4} \sim 10.2$\,K, are in good agreement with those of previous 
studies~\cite{nakamura_unique_2014,nakamura_transport_2015}. The inset in 
Fig.~\ref{fig:Tdepen}(a) shows a Curie-Weiss fit to the inverse 
susceptibility (for $T > 20$\,K), which yields an effective magnetic 
moment $\mu_\mathrm{eff} \sim 7.77$\,$\mu_\mathrm{B}$ and a paramagnetic 
Curie temperature $\theta_\mathrm{p} \sim 14.5$\,K. 
The effective moment is close to the theoretical value for free 
Eu$^{2+}$ ions (7.94\,$\mu_\mathrm{B}$). 
The AFM transitions can also be identified in the tem\-pe\-ra\-ture\--dependent 
$\rho_{xx}(T)$ data, yet they become less visible in an applied magnetic 
field. By contrast, the transitions are more evident in the field-dependent 
resistivity $\rho_{xx}(H)$ (see below). The high-$T$ resistivity data
are shown in the inset of Fig.~\ref{fig:Tdepen}(b). Here, the distinct
anomaly at $T_\mathrm{CDW} \sim 140$\,K is attributed to the gap opening
near a CDW transition~\cite{nakamura_magnetic_2013,araki_charge_2013,nakamura_unique_2014,nakamura_transport_2015,shimomura_lattice_2019,Kobata2016},
yet a more direct evidence is still missing.
Another notable feature in Fig.~\ref{fig:Tdepen}(b) is the giant MR at
base temperature, reaching $\sim 800$\,\% at 9\,T.
Since similar MR values have been reported also in nonmagnetic
BaAl$_4$~\cite{wang_crystalline_2020}, 
the magnetic nature of Eu$^{2+}$ ions cannot account for 
the appearance of mag\-ne\-to\-re\-sis\-tance in EuAl$_4$.

\begin{figure}
	\includegraphics[width=0.49\textwidth]{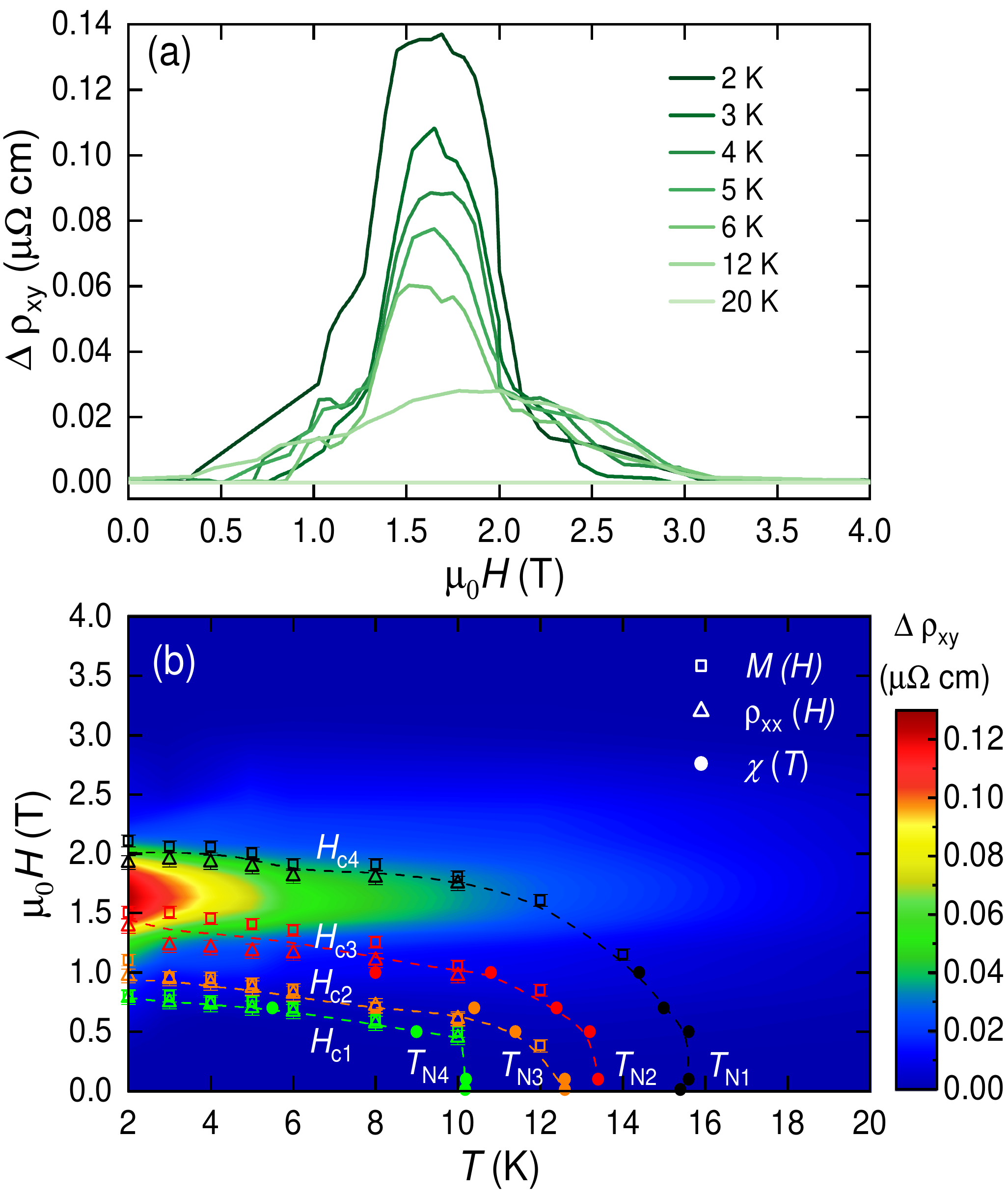}
	\centering
	\vspace{-2ex}%
	\caption{\label{fig:phase}(a) Field dependence of the extracted EuAl$_4$ 
	Hall resistivity $\Delta\rho_{xy}(H)$ at various temperatures (see text 
	for the definition of $\Delta\rho_{xy}$). 
	(b) Magnetic phase diagram of an EuAl$_4$, with the 
	field applied along the $c$-axis. The critical temperatures ($T_\mathrm{N1}$ 
	to $T_\mathrm{N4}$) are determined from $\chi(T,H)$ (circles), while 
	the critical fields are determined from $M(H,T)$ (squares) and 
	$\rho_{xx}(H,T)$ (triangles). The background color in (b) represents 
	the magnitude of $\Delta\rho_{xy}(H)$ at various temperatures. 
	The dashed lines are guides to the eyes. The error bars correspond to the field steps in the field-swept measurements.} 
\end{figure}

\begin{figure}
	\includegraphics[width=0.45\textwidth]{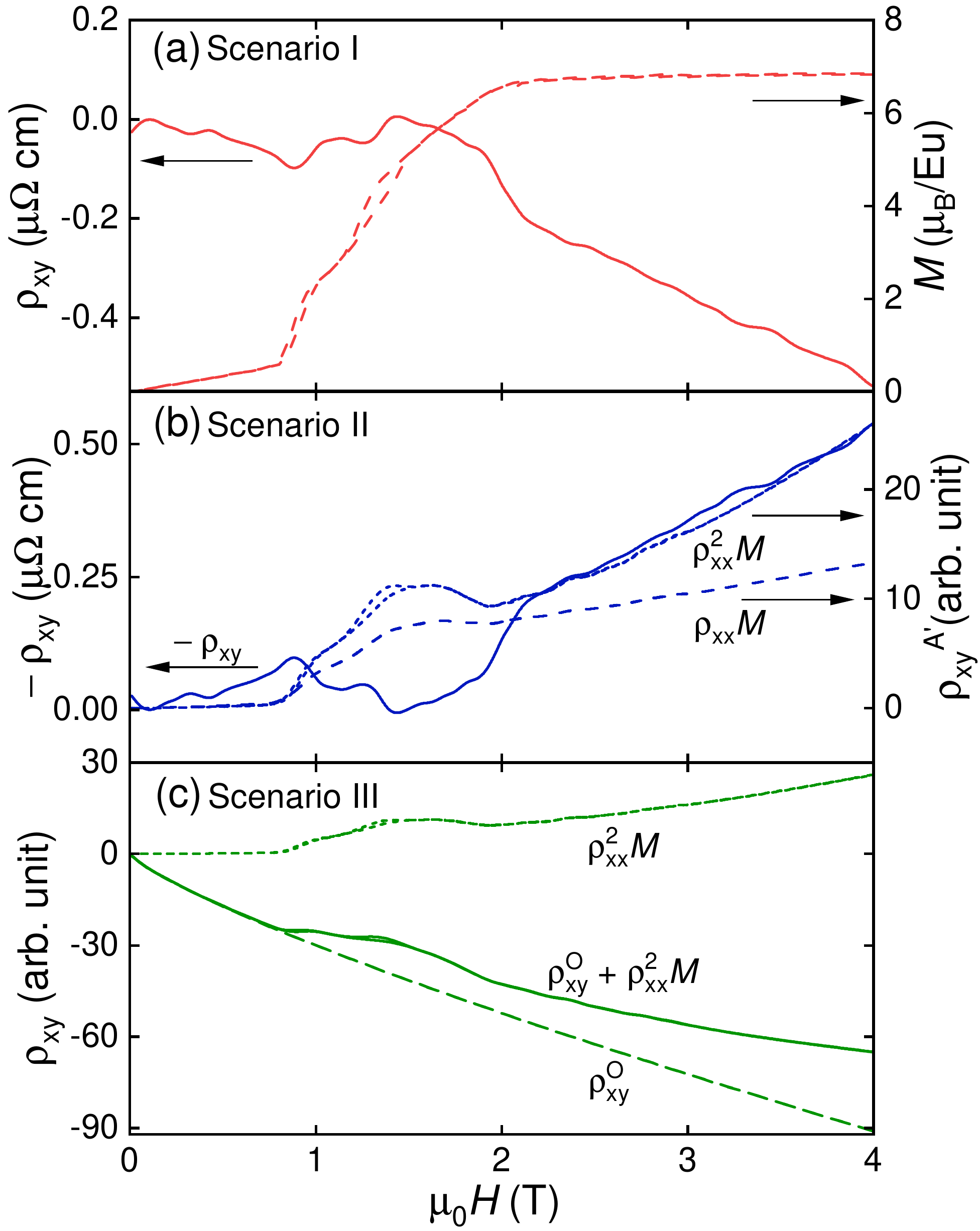}
	\centering
	\vspace{-2ex}%
	\caption{\label{fig:analysis}Possible scenarios for extracting the 
	topological contribution from the measured Hall resistivity of an 
	EuAl$_4$ single crystal.
	(a) Field-dependent Hall resistivity and magnetization. (b) Calculated 
	$\rho_{xy}^\mathrm{A'}(H)$. To compare it with the measured $\rho_{xy}(H)$, a scaling factor was used for calculating $\rho_{xy}^\mathrm{A'}$ from $\rho_{xx}$ and $M$. (c) Simulated field-dependent $\rho_{xy}$. The ordinary term is assumed to follow $\rho_{xy}^\mathrm{O} \propto H^{0.8}$.	 
		 All the results refer to the data at 2\,K.}
\end{figure}

Figure~\ref{fig:Hdepen} shows the field dependence of the magnetization
$M(H,T)$, electrical resistivity $\rho_{xx}(H,T)$, and Hall resistivity
$\rho_{xy}(H,T)$ of EuAl$_4$ at various temperatures, with the field
applied along the $c$-axis. The selected temperatures cover both the
antiferromagnetic and the paramagnetic states. In the AFM state (below 16\,K),
EuAl$_4$ undergoes three 
me\-ta\-mag\-ne\-tic transitions as the field increases.
At each transition, $M(H)$ shows a small yet clear hysteresis [see inset
in Fig.~\ref{fig:Hdepen}(a)].
By contrast, with the applied field in the $ab$-plane, the metamagnetic
transitions and the hysteresis are both less evident (see dashed line). 
At base temperature, the magnetization saturates when the external
field is larger than $\mu_0H_\mathrm{c4} \sim 2.1$\,T (red symbols).
For both field orientations, the saturation magnetization
$M_s \sim 6.8\,\mu_\mathrm{B}$ is consistent with 7.0\,$\mu_\mathrm{B}$,
the expected value for the $J = 7/2$ Eu$^{2+}$ ions.
For $H < H_\mathrm{c4}$, as indicated by arrows in Fig.~\ref{fig:Hdepen}(a),
EuAl$_4$ undergoes three metamagnetic transitions at $\mu_0H_\mathrm{c1} \sim 0.8$\,T,
$\mu_0H_\mathrm{c2} \sim 1.1$, and $\mu_0H_\mathrm{c3} \sim 1.5$,
respectively. The metamagnetic transitions are tracked also in the 
$\rho_{xx}(H)$ data. 
All the
critical fields, as determined from $\rho_{xx}(H,T)$, are highly consistent
with the magnetization results (see phase diagram below).

In the AFM state, in the field range between $H_\mathrm{c1}$ (first 
metamagnetic transition) and $H_\mathrm{c4}$ (saturation of magnetization), 
$\rho_{xy}(H,T)$ exhibits a hump-like anomaly [see Fig.~\ref{fig:Hdepen}(c)],  
reminiscent of the topological Hall resistivity arising from topological spin textures~\cite{neubauer_topological_2009,gayles_dzyaloshinskii-moriya_2015,kanazawa_large_2011,franz_real-space_2014,schulz_emergent_2012,qin_emergence_2019,matsuno_interface-driven_2016,kurumaji_skyrmion_2019,kanazawa_critical_2016,fujishiro_topological_2019,gobel_topological_2020,vistoli_giant_2019}.
The anomaly, particularly evident at low temperatures, becomes almost 
invisible above 12\,K. In general, to determine
the topological contribution $\rho_{xy}^T$, the
ordinary ($\rho_{xy}^O$) and the conventional anomalous ($\rho_{xy}^{A'}$) contributions 
have to be subtracted from the measured $\rho_{xy}$.
In EuAl$_4$, owing to its giant MR and the multiband origin of its 
ordinary Hall resistivity, such procedure is not feasible. 
The multiband nature of $\rho_{xy}$ is clearly evident from its nonlinear behavior for $\mu_0H$ > 6\,T [see inset of Fig.~\ref{fig:Hdepen}(c) and Ref.~\onlinecite{araki_charge_2013}].
This becomes even more robust upon 
applying a magnetic field 
in the $ab$-plane, thus making the subtraction of $\rho_{xy}^O$ 
unreliable.  
We recall that also the nonmagnetic BaAl$_4$ shows a multiband Hall 
resistivity in a wide temperature range~\cite{wang_crystalline_2020}.

Since the hump-like Hall resistivity appears only in a narrow field range, 
to extract the hump anomaly $\Delta\rho_{xy}(H)$ we may simply subtract a 
polynomial background [see black
line in the inset of Fig.~\ref{fig:Hdepen}(c)]. Note that $\Delta\rho_{xy}$ is part of the 
anomalous Hall resistivity, i.e., it might be either trivial (conventional 
anomalous Hall resistivity $\rho_{xy}^{A'}$) or nontrivial (topological 
Hall resistivity $\rho_{xy}^T$).
Independent of its nature, the derived $\Delta\rho_{xy}(H)$ at 
different temperatures are shown in Fig.~\ref{fig:phase}(a) and 
\ref{fig:phase}(b) (the latter as a contour plot). Clearly,
$\Delta\rho_{xy}$ is most prominent 
at temperatures below $T_\mathrm{N3}$ and
in the field range between $H_\mathrm{c3}$ and $H_\mathrm{c4}$, 
where the Eu$^{2+}$-moments undergo a third metamagnetic transition 
and become fully polarized. 

Now we discuss the different methods to decompose the measured Hall 
resistivity and hence track the
origin of its hump-like anomaly. 
To check whether a nonzero 
$\rho_{xy}^T$ underlies the hump in $\rho_{xy}(H)$, a knowledge of the 
exact field evolution of the ordinary-
$\rho_{xy}^O(H)$ and conventional anomalous Hall contributions 
$\rho_{xy}^{A'}(H)$ is crucial. 
On the one hand, as a compensated metal
\cite{[{A metal whose electron- and hole densities are equal. See, e.g.,}]Fawcett_1963,araki_charge_2013}, 
EuAl$_4$ exhibits multiple bands crossing the Fermi level, 
as confirmed experimentally by de Haas-van Alphen and photoelectron spectroscopy, 
and theoretically by band structure calculations~\cite{nakamura_unique_2014,nakamura_transport_2015,Kobata2016}. 
The ordinary Hall resistivity was previously analyzed using a two-carrier 
model in the paramagnetic state of EuAl$_4$~\cite{araki_charge_2013}. 
In the AFM state, such model becomes unreliable due to the presence of 
anomalous contributions. In this case, $\rho_{xy}^O(H)$ is unknown a priori, but it is presumably a nonlinear function of field.
On the other hand, the conventional anomalous Hall resistivity $\rho_{xy}^{A'}$ 
is even more complex to extract from the data. Initially, $\rho_{xy}^{A'}(H)$ 
was evaluated as $R_sM(H)$, with $R_s$ a constant and $M(H)$ the 
field-dependent magnetization~\cite{nagaosa_anomalous_2010}. 
Later on it was recognized that the coefficient $R_s$ is not a constant, 
but rather a function of the field-dependent longitudinal electrical 
resistivity $\rho_{xx}(H)$~\cite{lee_hidden_2007}. Consequently, 
$\rho_{xy}^{A'}$ can be rewritten as $S_H\rho_{xx}^2M$ or 
$S_H'\rho_{xx}M$. In real materials, $\rho_{xy}^{A'}$ depends on the 
mechanisms of intrinsic-, side-jump, or skew scattering, or an 
intricate combination thereof~\cite{nagaosa_anomalous_2010,tian_proper_2009,hou_multivariable_2015}.
These different representations together with the multiband nature of 
EuAl$_4$ make the extraction of $\rho_{xy}^T$ from the measured 
$\rho_{xy}$ even more complicated, especially considering the presence 
of a giant MR (implying a large $\rho_{xx}$) in EuAl$_4$. 
Below we discuss in detail three possible scenarios.

Scenario I: \emph{$\rho_{xy}^{A'}$ proportional to $M$.} Despite 
its simplicity, 
this scenario has often been used, especially in ferromagnets~\cite{Ying2012,nagaosa_anomalous_2010}. 
The $M(H)$ data in Fig.~\ref{fig:analysis}(a) show that the 
magnetization of EuAl$_4$ undergoes step-like metamagnetic transitions, 
to finally saturate above 2.1\,T. Consequently, in principle, $\rho_{xy}^{A'}$ 
should exhibit similar features to the magnetization. However, 
instead of a step-like feature, a hump-like anomaly is observed in 
the Hall resistivity of EuAl$_4$. Clearly, if this scenario 
applies, $\rho_{xy}^{A'}$ contributes negligibly to the total $\rho_{xy}$. 
This implies that the hump-like anomaly is basically due to the topological 
$\rho_{xy}^T$ term, most likely caused by topological spin textures.	    

Scenario II: \emph{$\rho_{xy}^{A'}$ proportional to $\rho_{xx}^2 M$ or 
$\rho_{xx} M$ with a negative prefactor.} If the prefactors $S_H$ and $S_H'$ 
are allowed to be negative, then the hump anomaly in $\rho_{xx}^2M(H)$ 
or $\rho_{xx}M(H)$ is convex [see Fig.~\ref{fig:analysis}(b)], which is 
opposite to the concave feature in $-\rho_{xy}(H)$. Since both $M(H)$ and 
$\rho_{xx}{}(H)$ are positive, a negative sign is necessary to overlap 
the calculated $\rho_{xy}^{A'}$ with the measured $\rho_{xy}$, which 
is negative at high magnetic fields [see $\rho_{xy}(H)$ in Fig.~\ref{fig:analysis}(a)].
In this case, a subtraction of $\rho_{xy}^{A'}$ from $\rho_{xy}$ 
enhances the hump feature, 
i.e., it increases the magnitude of $\Delta\rho_{xy}$. Again, such 
scenario implies that the 
anomaly in $\rho_{xy}$ must come 
from a topological term $\rho_{xy}^T$, and the extracted $\Delta\rho_{xy}$ 
shown in Fig.~\ref{fig:phase} represents a lower limit to the 
intrinsic value of $\rho_{xy}^T$. This scenario was successfully 
applied to extract the $\rho_{xy}^T$ in EuPtSi, whose $A$-phase 
is proposed to host magnetic skyrmions~\cite{kakihana_giant_2017}.

Scenario III: \emph{$\rho_{xy}^{A'}$ proportional to $\rho_{xx}^2 M$ 
or $\rho_{xx} M$ with a positive prefactor.}
If the prefactors $S_H$ and $S_H'$ are allowed to 
be positive, then the outcome 
could be different.
We simulated the behavior of $\rho_{xy}(H)$ by combining $\rho_{xy}^O$ 
and $\rho_{xx}^2M$, assuming that $\rho_{xy}^O(H)$ follows an $H^{0.8}$ 
dependence.
As shown by the solid line in Fig.~\ref{fig:analysis}(c), the simulated 
$\rho_{xy}(H)$ qualitatively agrees with the measured $\rho_{xy}(H)$. 
In this case, the hump anomaly in $\rho_{xy}$ is trivial, being closely 
related to $\rho_{xy}^{A'}$, and no additional $\rho_{xy}^T$ contribution 
needs to be invoked. 
However, even in such case, a finite $\rho_{xy}^T$ might still exist,  
although mostly masked by $\rho_{xy}^{A'}$. This underlying topological 
component, though difficult to isolate, can still contribute to the hump 
anomaly in $\rho_{xy}$, as shown, e.g., in EuCd$_2$As$_2$ and 
CeAlGe~\cite{xu_topological_2020,puphal_topological_2020}. 
In this case, the extracted $\Delta\rho_{xy}$ shown in Fig.~\ref{fig:phase} 
represents an upper limit to the intrinsic value of $\rho_{xy}^T$.

Depending on which scenario applies, the interpretation of Hall 
resistivity data of EuAl$_4$ is different. Now we further discuss the 
nontrivial origin of the hump-like anomaly in $\rho_{xy}$. 
The observation of a
topological Hall effect is usually attributed to 
noncoplanar spin textures, such as magnetic skyrmions, characterized by 
a finite scalar spin chirality in real space. 
These spin textures are often observed in magnetic materials 
that lack an inversion symmetry, 
and can be stabilized by the Dzyaloshinskii-Moriya interaction~\cite{muhlbauer_skyrmion_2009,yu_near_2011,yu_real-space_2010,seki_observation_2012,kezsmarki_ne-type_2015,tokunaga_new_2015,Seki2012}. 
Conversely, magnetic materials with a centrosymmetric crystal 
structure that still host magnetic skyrmions are rare. To date, 
only a few systems have been reported, including 
Gd$_2$PdSi$_3$~\cite{kurumaji_skyrmion_2019}, Gd$_3$Ru$_{4}$Al$_{12}$~\cite{Hirschberger2019}, Fe$_3$Sn$_2$~\cite{li_large_2019}, 
and recently GdRu$_2$Si$_2$~\cite{Khanh2020}. 
Compared to noncentrosymmetric systems, skyrmions in centrosymmetric 
materials exhibit the unique advantages of tunable skyrmion size and spin 
helicity~\cite{yu_biskyrmion_2014}. In centrosymmetric systems, for 
example, skyrmions can be stabilized either by magnetic frustration 
(e.g., in Gd$_3$Ru$_{4}$Al$_{12}$, Gd$_2$PdSi$_3$, and Fe$_3$Sn$_2$), 
or by the competition between the magnetic interactions and magnetic 
anisotropies (e.g., in GdRu$_2$Si$_2$)~\cite{Batista2016,Hirschberger2019,kurumaji_skyrmion_2019,li_large_2019,Khanh2020}. 
In the EuAl$_4$ case, as shown in the inset of Fig.~\ref{fig:Hdepen}(a), 
the magnetic anisotropy is moderate. Yet, according to recent NMR studies, 
EuAl$_4$ exhibits a clear anisotropic Knight shift as the temperature 
approaches $T_\mathrm{N}$~\cite{Niki2015}. Since EuAl$_4$ adopts the 
same crystal structure of GdRu$_2$Si$_2$, skyrmions might be stabilized 
by the same mechanism. More interestingly, if topological spin textures 
indeed exist in the AFM phase of EuAl$_4$, 
this would represent a rare case where a rather exotic magnetic order 
coexists with CDW order. 

Apart from the above topological spin textures, upon breaking certain 
symmetries, noncollinear antiferromagnets 
may also exhibit a topological Hall effect due to crossings or anticrossings 
of bands with a significant Berry curvature, as e.g., at the Weyl points~\cite{THEnote}. 
Such a momentum-space scenario has been theoretically proposed and 
experimentally observed, for instance, in Mn$_3$Sn~\cite{nakatsuji_large_2015,ikhlas_large_2017}, 
GdPtBi~\cite{suzuki_large_2016}, YbPtBi~\cite{guo_evidence_2018}, 
and Mn$_3$Ge~\cite{nayak_large_2016}. 
A three-dimensional Dirac spectrum with nontrivial topology and possible 
nodal-lines crossing the Brillouin zone was recently observed in nonmagnetic BaAl$_4$~\cite{wang_crystalline_2020}.
In this context, a topologically nontrivial band structure is also 
expected in magnetic EuAl$_4$, extending beyond 
its magnetically ordered state.

In summary, we observed a hump-like anomaly $\Delta\rho_{xy}$ in the 
Hall resistivity of the centrosymmetric antiferromagnet EuAl$_4$ 
(single crystal). By systematic field- and temperature-dependent 
electrical resistivity and magnetization measurements, we could 
establish the magnetic phase diagram of EuAl$_4$. 
The $\Delta\rho_{xy}$ anomaly appears mostly in a field range 
where also metamagnetic transitions occur. 
Depending on the scenario used for evaluating the conventional anomalous 
Hall resistivity, the observed $\Delta\rho_{xy}$ corresponds to a 
topological Hall term $\rho_{xy}^T$, or to the lower/upper limits of 
the topological contribution. Although a trivial origin of the effect 
cannot be fully excluded, our results suggest that a topological Hall 
effect and topological spin textures may indeed exist in EuAl$_4$. 
To confirm such topological magnetic phase in EuAl$_4$, further experiments, 
as resonant x-ray scattering or Lorentz transmission electron microscopy, 
are highly desirable.
EuAl$_4$ represents a rare case where both geometrical frustration and 
inversion symmetry breaking are absent. Hence, it may offer a candidate 
compound for exploring the skyrmion physics and its applications in 
materials with a simple crystal structure.


We thank M.\ Medarde for the assistance during the electrical 
resistivity and magnetization measurements. T.\ Shang acknowledges 
support from the Schwei\-ze\-rische Na\-ti\-o\-nal\-fonds zur F\"{o}r\-de\-rung
der Wis\-sen\-schaft\-lich\-en For\-schung, SNF (Grants No.\ 200021\_188706 
and 206021\_139082). Y.\ Xu was supported by SNF via Grants No.\ 206021\_139082 
and PP00P2\_179097. This work was partially supported also by the National 
Natural Science Foundation of China (Grants No.\ 11674336 and 11874150) and the Sino-Swiss Science and Technology Cooperation (Grant No. IZLCZ2-170075).



\bibliography{EuAl4}

\begin{thebibliography}{52}%
\makeatletter
\providecommand \@ifxundefined [1]{%
 \@ifx{#1\undefined}
}%
\providecommand \@ifnum [1]{%
 \ifnum #1\expandafter \@firstoftwo
 \else \expandafter \@secondoftwo
 \fi
}%
\providecommand \@ifx [1]{%
 \ifx #1\expandafter \@firstoftwo
 \else \expandafter \@secondoftwo
 \fi
}%
\providecommand \natexlab [1]{#1}%
\providecommand \enquote  [1]{``#1''}%
\providecommand \bibnamefont  [1]{#1}%
\providecommand \bibfnamefont [1]{#1}%
\providecommand \citenamefont [1]{#1}%
\providecommand \href@noop [0]{\@secondoftwo}%
\providecommand \href [0]{\begingroup \@sanitize@url \@href}%
\providecommand \@href[1]{\@@startlink{#1}\@@href}%
\providecommand \@@href[1]{\endgroup#1\@@endlink}%
\providecommand \@sanitize@url [0]{\catcode `\\12\catcode `\$12\catcode
  `\&12\catcode `\#12\catcode `\^12\catcode `\_12\catcode `\%12\relax}%
\providecommand \@@startlink[1]{}%
\providecommand \@@endlink[0]{}%
\providecommand \url  [0]{\begingroup\@sanitize@url \@url }%
\providecommand \@url [1]{\endgroup\@href {#1}{\urlprefix }}%
\providecommand \urlprefix  [0]{URL }%
\providecommand \Eprint [0]{\href }%
\providecommand \doibase [0]{https://doi.org/}%
\providecommand \selectlanguage [0]{\@gobble}%
\providecommand \bibinfo  [0]{\@secondoftwo}%
\providecommand \bibfield  [0]{\@secondoftwo}%
\providecommand \translation [1]{[#1]}%
\providecommand \BibitemOpen [0]{}%
\providecommand \bibitemStop [0]{}%
\providecommand \bibitemNoStop [0]{.\EOS\space}%
\providecommand \EOS [0]{\spacefactor3000\relax}%
\providecommand \BibitemShut  [1]{\csname bibitem#1\endcsname}%
\let\auto@bib@innerbib\@empty
\bibitem [{\citenamefont {Nagaosa}\ and\ \citenamefont
  {Tokura}(2013)}]{Nagaosa2013}%
  \BibitemOpen
  \bibfield  {author} {\bibinfo {author} {\bibfnamefont {N.}~\bibnamefont
  {Nagaosa}}\ and\ \bibinfo {author} {\bibfnamefont {Y.}~\bibnamefont
  {Tokura}},\ }\bibfield  {title} {\bibinfo {title} {Topological properties and
  dynamics of magnetic skyrmions},\ }\href
  {https://doi.org/10.1038/nnano.2013.243} {\bibfield  {journal} {\bibinfo
  {journal} {Nat. Nanotechnol.}\ }\textbf {\bibinfo {volume} {8}},\ \bibinfo
  {pages} {899} (\bibinfo {year} {2013})}\BibitemShut {NoStop}%
\bibitem [{\citenamefont {Sinova}\ \emph {et~al.}(2015)\citenamefont {Sinova},
  \citenamefont {Valenzuela}, \citenamefont {Wunderlich}, \citenamefont
  {Back},\ and\ \citenamefont {Jungwirth}}]{Sinova2015}%
  \BibitemOpen
  \bibfield  {author} {\bibinfo {author} {\bibfnamefont {J.}~\bibnamefont
  {Sinova}}, \bibinfo {author} {\bibfnamefont {S.~O.}\ \bibnamefont
  {Valenzuela}}, \bibinfo {author} {\bibfnamefont {J.}~\bibnamefont
  {Wunderlich}}, \bibinfo {author} {\bibfnamefont {C.~H.}\ \bibnamefont
  {Back}},\ and\ \bibinfo {author} {\bibfnamefont {T.}~\bibnamefont
  {Jungwirth}},\ }\bibfield  {title} {\bibinfo {title} {{Spin} {Hall}
  effects},\ }\href {https://doi.org/10.1103/RevModPhys.87.1213} {\bibfield
  {journal} {\bibinfo  {journal} {Rev. Mod. Phys.}\ }\textbf {\bibinfo {volume}
  {87}},\ \bibinfo {pages} {1213} (\bibinfo {year} {2015})}\BibitemShut
  {NoStop}%
\bibitem [{\citenamefont {Hirohata}\ \emph {et~al.}(2020)\citenamefont
  {Hirohata}, \citenamefont {Yamada}, \citenamefont {Nakatani}, \citenamefont
  {Prejbeanu}, \citenamefont {Di\'{e}ny}, \citenamefont {Pirro},\ and\
  \citenamefont {Hillebrands}}]{Hirohata2020}%
  \BibitemOpen
  \bibfield  {author} {\bibinfo {author} {\bibfnamefont {A.}~\bibnamefont
  {Hirohata}}, \bibinfo {author} {\bibfnamefont {K.}~\bibnamefont {Yamada}},
  \bibinfo {author} {\bibfnamefont {Y.}~\bibnamefont {Nakatani}}, \bibinfo
  {author} {\bibfnamefont {I.-L.}\ \bibnamefont {Prejbeanu}}, \bibinfo {author}
  {\bibfnamefont {B.}~\bibnamefont {Di\'{e}ny}}, \bibinfo {author}
  {\bibfnamefont {P.}~\bibnamefont {Pirro}},\ and\ \bibinfo {author}
  {\bibfnamefont {B.}~\bibnamefont {Hillebrands}},\ }\bibfield  {title}
  {\bibinfo {title} {Review on spintronics: {P}rinciples and device
  applications},\ }\href {https://doi.org/10.1016/j.jmmm.2020.166711}
  {\bibfield  {journal} {\bibinfo  {journal} {J. Magn. Magn. Mater.}\ }\textbf
  {\bibinfo {volume} {509}},\ \bibinfo {pages} {166711} (\bibinfo {year}
  {2020})}\BibitemShut {NoStop}%
\bibitem [{\citenamefont {Neubauer}\ \emph {et~al.}(2009)\citenamefont
  {Neubauer}, \citenamefont {Pfleiderer}, \citenamefont {Binz}, \citenamefont
  {Rosch}, \citenamefont {Ritz}, \citenamefont {Niklowitz},\ and\ \citenamefont
  {B\"oni}}]{neubauer_topological_2009}%
  \BibitemOpen
  \bibfield  {author} {\bibinfo {author} {\bibfnamefont {A.}~\bibnamefont
  {Neubauer}}, \bibinfo {author} {\bibfnamefont {C.}~\bibnamefont
  {Pfleiderer}}, \bibinfo {author} {\bibfnamefont {B.}~\bibnamefont {Binz}},
  \bibinfo {author} {\bibfnamefont {A.}~\bibnamefont {Rosch}}, \bibinfo
  {author} {\bibfnamefont {R.}~\bibnamefont {Ritz}}, \bibinfo {author}
  {\bibfnamefont {P.~G.}\ \bibnamefont {Niklowitz}},\ and\ \bibinfo {author}
  {\bibfnamefont {P.}~\bibnamefont {B\"oni}},\ }\bibfield  {title} {\bibinfo
  {title} {Topological {Hall} effect in the {$A$} phase of {MnSi}},\ }\href
  {https://doi.org/10.1103/PhysRevLett.102.186602} {\bibfield  {journal}
  {\bibinfo  {journal} {Phys. Rev. Lett.}\ }\textbf {\bibinfo {volume} {102}},\
  \bibinfo {pages} {186602} (\bibinfo {year} {2009})}\BibitemShut {NoStop}%
\bibitem [{\citenamefont {Gayles}\ \emph {et~al.}(2015)\citenamefont {Gayles},
  \citenamefont {Freimuth}, \citenamefont {Schena}, \citenamefont {Lani},
  \citenamefont {Mavropoulos}, \citenamefont {Duine}, \citenamefont {Bl\"ugel},
  \citenamefont {Sinova},\ and\ \citenamefont
  {Mokrousov}}]{gayles_dzyaloshinskii-moriya_2015}%
  \BibitemOpen
  \bibfield  {author} {\bibinfo {author} {\bibfnamefont {J.}~\bibnamefont
  {Gayles}}, \bibinfo {author} {\bibfnamefont {F.}~\bibnamefont {Freimuth}},
  \bibinfo {author} {\bibfnamefont {T.}~\bibnamefont {Schena}}, \bibinfo
  {author} {\bibfnamefont {G.}~\bibnamefont {Lani}}, \bibinfo {author}
  {\bibfnamefont {P.}~\bibnamefont {Mavropoulos}}, \bibinfo {author}
  {\bibfnamefont {R.~A.}\ \bibnamefont {Duine}}, \bibinfo {author}
  {\bibfnamefont {S.}~\bibnamefont {Bl\"ugel}}, \bibinfo {author}
  {\bibfnamefont {J.}~\bibnamefont {Sinova}},\ and\ \bibinfo {author}
  {\bibfnamefont {Y.}~\bibnamefont {Mokrousov}},\ }\bibfield  {title} {\bibinfo
  {title} {Dzya\-lo\-shin\-skii-{Moriya} interaction and {Hall} effects in the
  skyrmion phase of {Mn$_{1-x}$Fe$_x$Ge}},\ }\href
  {https://doi.org/10.1103/PhysRevLett.115.036602} {\bibfield  {journal}
  {\bibinfo  {journal} {Phys. Rev. Lett.}\ }\textbf {\bibinfo {volume} {115}},\
  \bibinfo {pages} {036602} (\bibinfo {year} {2015})}\BibitemShut {NoStop}%
\bibitem [{\citenamefont {Kanazawa}\ \emph {et~al.}(2011)\citenamefont
  {Kanazawa}, \citenamefont {Onose}, \citenamefont {Arima}, \citenamefont
  {Okuyama}, \citenamefont {Ohoyama}, \citenamefont {Wakimoto}, \citenamefont
  {Kakurai}, \citenamefont {Ishiwata},\ and\ \citenamefont
  {Tokura}}]{kanazawa_large_2011}%
  \BibitemOpen
  \bibfield  {author} {\bibinfo {author} {\bibfnamefont {N.}~\bibnamefont
  {Kanazawa}}, \bibinfo {author} {\bibfnamefont {Y.}~\bibnamefont {Onose}},
  \bibinfo {author} {\bibfnamefont {T.}~\bibnamefont {Arima}}, \bibinfo
  {author} {\bibfnamefont {D.}~\bibnamefont {Okuyama}}, \bibinfo {author}
  {\bibfnamefont {K.}~\bibnamefont {Ohoyama}}, \bibinfo {author} {\bibfnamefont
  {S.}~\bibnamefont {Wakimoto}}, \bibinfo {author} {\bibfnamefont
  {K.}~\bibnamefont {Kakurai}}, \bibinfo {author} {\bibfnamefont
  {S.}~\bibnamefont {Ishiwata}},\ and\ \bibinfo {author} {\bibfnamefont
  {Y.}~\bibnamefont {Tokura}},\ }\bibfield  {title} {\bibinfo {title} {Large
  topological {Hall} effect in a short-period helimagnet {MnGe}},\ }\href
  {https://doi.org/10.1103/PhysRevLett.106.156603} {\bibfield  {journal}
  {\bibinfo  {journal} {Phys. Rev. Lett.}\ }\textbf {\bibinfo {volume} {106}},\
  \bibinfo {pages} {156603} (\bibinfo {year} {2011})}\BibitemShut {NoStop}%
\bibitem [{\citenamefont {Franz}\ \emph {et~al.}(2014)\citenamefont {Franz},
  \citenamefont {Freimuth}, \citenamefont {Bauer}, \citenamefont {Ritz},
  \citenamefont {Schnarr}, \citenamefont {Duvinage}, \citenamefont {Adams},
  \citenamefont {Bl\"ugel}, \citenamefont {Rosch}, \citenamefont {Mokrousov},\
  and\ \citenamefont {Pfleiderer}}]{franz_real-space_2014}%
  \BibitemOpen
  \bibfield  {author} {\bibinfo {author} {\bibfnamefont {C.}~\bibnamefont
  {Franz}}, \bibinfo {author} {\bibfnamefont {F.}~\bibnamefont {Freimuth}},
  \bibinfo {author} {\bibfnamefont {A.}~\bibnamefont {Bauer}}, \bibinfo
  {author} {\bibfnamefont {R.}~\bibnamefont {Ritz}}, \bibinfo {author}
  {\bibfnamefont {C.}~\bibnamefont {Schnarr}}, \bibinfo {author} {\bibfnamefont
  {C.}~\bibnamefont {Duvinage}}, \bibinfo {author} {\bibfnamefont
  {T.}~\bibnamefont {Adams}}, \bibinfo {author} {\bibfnamefont
  {S.}~\bibnamefont {Bl\"ugel}}, \bibinfo {author} {\bibfnamefont
  {A.}~\bibnamefont {Rosch}}, \bibinfo {author} {\bibfnamefont
  {Y.}~\bibnamefont {Mokrousov}},\ and\ \bibinfo {author} {\bibfnamefont
  {C.}~\bibnamefont {Pfleiderer}},\ }\bibfield  {title} {\bibinfo {title}
  {Real-space and reciprocal-space {Berry} phases in the {Hall} effect of
  {Mn$_{1-x}$Fe$_x$Si}},\ }\href
  {https://doi.org/10.1103/PhysRevLett.112.186601} {\bibfield  {journal}
  {\bibinfo  {journal} {Phys. Rev. Lett.}\ }\textbf {\bibinfo {volume} {112}},\
  \bibinfo {pages} {186601} (\bibinfo {year} {2014})}\BibitemShut {NoStop}%
\bibitem [{\citenamefont {Schulz}\ \emph {et~al.}(2012)\citenamefont {Schulz},
  \citenamefont {Ritz}, \citenamefont {Bauer}, \citenamefont {Halder},
  \citenamefont {Wagner}, \citenamefont {Franz}, \citenamefont {Pfleiderer},
  \citenamefont {Everschor}, \citenamefont {Garst},\ and\ \citenamefont
  {Rosch}}]{schulz_emergent_2012}%
  \BibitemOpen
  \bibfield  {author} {\bibinfo {author} {\bibfnamefont {T.}~\bibnamefont
  {Schulz}}, \bibinfo {author} {\bibfnamefont {R.}~\bibnamefont {Ritz}},
  \bibinfo {author} {\bibfnamefont {A.}~\bibnamefont {Bauer}}, \bibinfo
  {author} {\bibfnamefont {M.}~\bibnamefont {Halder}}, \bibinfo {author}
  {\bibfnamefont {M.}~\bibnamefont {Wagner}}, \bibinfo {author} {\bibfnamefont
  {C.}~\bibnamefont {Franz}}, \bibinfo {author} {\bibfnamefont
  {C.}~\bibnamefont {Pfleiderer}}, \bibinfo {author} {\bibfnamefont
  {K.}~\bibnamefont {Everschor}}, \bibinfo {author} {\bibfnamefont
  {M.}~\bibnamefont {Garst}},\ and\ \bibinfo {author} {\bibfnamefont
  {A.}~\bibnamefont {Rosch}},\ }\bibfield  {title} {\bibinfo {title} {Emergent
  electrodynamics of skyrmions in a chiral magnet},\ }\href
  {https://doi.org/10.1038/nphys2231} {\bibfield  {journal} {\bibinfo
  {journal} {Nat. Phys.}\ }\textbf {\bibinfo {volume} {8}},\ \bibinfo {pages}
  {301} (\bibinfo {year} {2012})}\BibitemShut {NoStop}%
\bibitem [{\citenamefont {Qin}\ \emph {et~al.}(2019)\citenamefont {Qin},
  \citenamefont {Liu}, \citenamefont {Lin}, \citenamefont {Shu}, \citenamefont
  {Xie}, \citenamefont {Lim}, \citenamefont {Li}, \citenamefont {He},
  \citenamefont {Chow},\ and\ \citenamefont {Chen}}]{qin_emergence_2019}%
  \BibitemOpen
  \bibfield  {author} {\bibinfo {author} {\bibfnamefont {Q.}~\bibnamefont
  {Qin}}, \bibinfo {author} {\bibfnamefont {L.}~\bibnamefont {Liu}}, \bibinfo
  {author} {\bibfnamefont {W.}~\bibnamefont {Lin}}, \bibinfo {author}
  {\bibfnamefont {X.}~\bibnamefont {Shu}}, \bibinfo {author} {\bibfnamefont
  {Q.}~\bibnamefont {Xie}}, \bibinfo {author} {\bibfnamefont {Z.}~\bibnamefont
  {Lim}}, \bibinfo {author} {\bibfnamefont {C.}~\bibnamefont {Li}}, \bibinfo
  {author} {\bibfnamefont {S.}~\bibnamefont {He}}, \bibinfo {author}
  {\bibfnamefont {G.~M.}\ \bibnamefont {Chow}},\ and\ \bibinfo {author}
  {\bibfnamefont {J.}~\bibnamefont {Chen}},\ }\bibfield  {title} {\bibinfo
  {title} {Emergence of topological {Hall} effect in a {SrRuO$_3$} single
  layer},\ }\href {https://doi.org/10.1002/adma.201807008} {\bibfield
  {journal} {\bibinfo  {journal} {Adv. Mater.}\ }\textbf {\bibinfo {volume}
  {31}},\ \bibinfo {pages} {1807008} (\bibinfo {year} {2019})}\BibitemShut
  {NoStop}%
\bibitem [{\citenamefont {Matsuno}\ \emph {et~al.}(2016)\citenamefont
  {Matsuno}, \citenamefont {Ogawa}, \citenamefont {Yasuda}, \citenamefont
  {Kagawa}, \citenamefont {Koshibae}, \citenamefont {Nagaosa}, \citenamefont
  {Tokura},\ and\ \citenamefont {Kawasaki}}]{matsuno_interface-driven_2016}%
  \BibitemOpen
  \bibfield  {author} {\bibinfo {author} {\bibfnamefont {J.}~\bibnamefont
  {Matsuno}}, \bibinfo {author} {\bibfnamefont {N.}~\bibnamefont {Ogawa}},
  \bibinfo {author} {\bibfnamefont {K.}~\bibnamefont {Yasuda}}, \bibinfo
  {author} {\bibfnamefont {F.}~\bibnamefont {Kagawa}}, \bibinfo {author}
  {\bibfnamefont {W.}~\bibnamefont {Koshibae}}, \bibinfo {author}
  {\bibfnamefont {N.}~\bibnamefont {Nagaosa}}, \bibinfo {author} {\bibfnamefont
  {Y.}~\bibnamefont {Tokura}},\ and\ \bibinfo {author} {\bibfnamefont
  {M.}~\bibnamefont {Kawasaki}},\ }\bibfield  {title} {\bibinfo {title}
  {Interface-driven topological {Hall} effect in {SrRuO$_3$}-{SrIrO$_3$}
  bilayer},\ }\href {https://doi.org/10.1126/sciadv.1600304} {\bibfield
  {journal} {\bibinfo  {journal} {Sci. Adv.}\ }\textbf {\bibinfo {volume}
  {2}},\ \bibinfo {pages} {e1600304} (\bibinfo {year} {2016})}\BibitemShut
  {NoStop}%
\bibitem [{\citenamefont {Kurumaji}\ \emph {et~al.}(2019)\citenamefont
  {Kurumaji}, \citenamefont {Nakajima}, \citenamefont {Hirschberger},
  \citenamefont {Kikkawa}, \citenamefont {Yamasaki}, \citenamefont {Sagayama},
  \citenamefont {Nakao}, \citenamefont {Taguchi}, \citenamefont {Arima},\ and\
  \citenamefont {Tokura}}]{kurumaji_skyrmion_2019}%
  \BibitemOpen
  \bibfield  {author} {\bibinfo {author} {\bibfnamefont {T.}~\bibnamefont
  {Kurumaji}}, \bibinfo {author} {\bibfnamefont {T.}~\bibnamefont {Nakajima}},
  \bibinfo {author} {\bibfnamefont {M.}~\bibnamefont {Hirschberger}}, \bibinfo
  {author} {\bibfnamefont {A.}~\bibnamefont {Kikkawa}}, \bibinfo {author}
  {\bibfnamefont {Y.}~\bibnamefont {Yamasaki}}, \bibinfo {author}
  {\bibfnamefont {H.}~\bibnamefont {Sagayama}}, \bibinfo {author}
  {\bibfnamefont {H.}~\bibnamefont {Nakao}}, \bibinfo {author} {\bibfnamefont
  {Y.}~\bibnamefont {Taguchi}}, \bibinfo {author} {\bibfnamefont {T.-h.}\
  \bibnamefont {Arima}},\ and\ \bibinfo {author} {\bibfnamefont
  {Y.}~\bibnamefont {Tokura}},\ }\bibfield  {title} {\bibinfo {title} {Skyrmion
  lattice with a giant topological {Hall} effect in a frustrated
  triangular-lattice magnet},\ }\href {https://doi.org/10.1126/science.aau0968}
  {\bibfield  {journal} {\bibinfo  {journal} {Science}\ }\textbf {\bibinfo
  {volume} {365}},\ \bibinfo {pages} {914} (\bibinfo {year}
  {2019})}\BibitemShut {NoStop}%
\bibitem [{\citenamefont {Kanazawa}\ \emph {et~al.}(2016)\citenamefont
  {Kanazawa}, \citenamefont {Nii}, \citenamefont {Zhang}, \citenamefont
  {Mishchenko}, \citenamefont {De~Filippis}, \citenamefont {Kagawa},
  \citenamefont {Iwasa}, \citenamefont {Nagaosa},\ and\ \citenamefont
  {Tokura}}]{kanazawa_critical_2016}%
  \BibitemOpen
  \bibfield  {author} {\bibinfo {author} {\bibfnamefont {N.}~\bibnamefont
  {Kanazawa}}, \bibinfo {author} {\bibfnamefont {Y.}~\bibnamefont {Nii}},
  \bibinfo {author} {\bibfnamefont {X.-X.}\ \bibnamefont {Zhang}}, \bibinfo
  {author} {\bibfnamefont {A.~S.}\ \bibnamefont {Mishchenko}}, \bibinfo
  {author} {\bibfnamefont {G.}~\bibnamefont {De~Filippis}}, \bibinfo {author}
  {\bibfnamefont {F.}~\bibnamefont {Kagawa}}, \bibinfo {author} {\bibfnamefont
  {Y.}~\bibnamefont {Iwasa}}, \bibinfo {author} {\bibfnamefont
  {N.}~\bibnamefont {Nagaosa}},\ and\ \bibinfo {author} {\bibfnamefont
  {Y.}~\bibnamefont {Tokura}},\ }\bibfield  {title} {\bibinfo {title} {Critical
  phenomena of emergent magnetic monopoles in a chiral magnet},\ }\href
  {https://doi.org/10.1038/ncomms11622} {\bibfield  {journal} {\bibinfo
  {journal} {Nat. Commun.}\ }\textbf {\bibinfo {volume} {7}},\ \bibinfo {pages}
  {1} (\bibinfo {year} {2016})}\BibitemShut {NoStop}%
\bibitem [{\citenamefont {Fujishiro}\ \emph {et~al.}(2019)\citenamefont
  {Fujishiro}, \citenamefont {Kanazawa}, \citenamefont {Nakajima},
  \citenamefont {Yu}, \citenamefont {Ohishi}, \citenamefont {Kawamura},
  \citenamefont {Kakurai}, \citenamefont {Arima}, \citenamefont {Mitamura},
  \citenamefont {Miyake}, \citenamefont {Akiba}, \citenamefont {Tokunaga},
  \citenamefont {Matsuo}, \citenamefont {Kindo}, \citenamefont {Koretsune},
  \citenamefont {Arita},\ and\ \citenamefont
  {Tokura}}]{fujishiro_topological_2019}%
  \BibitemOpen
  \bibfield  {author} {\bibinfo {author} {\bibfnamefont {Y.}~\bibnamefont
  {Fujishiro}}, \bibinfo {author} {\bibfnamefont {N.}~\bibnamefont {Kanazawa}},
  \bibinfo {author} {\bibfnamefont {T.}~\bibnamefont {Nakajima}}, \bibinfo
  {author} {\bibfnamefont {X.~Z.}\ \bibnamefont {Yu}}, \bibinfo {author}
  {\bibfnamefont {K.}~\bibnamefont {Ohishi}}, \bibinfo {author} {\bibfnamefont
  {Y.}~\bibnamefont {Kawamura}}, \bibinfo {author} {\bibfnamefont
  {K.}~\bibnamefont {Kakurai}}, \bibinfo {author} {\bibfnamefont
  {T.}~\bibnamefont {Arima}}, \bibinfo {author} {\bibfnamefont
  {H.}~\bibnamefont {Mitamura}}, \bibinfo {author} {\bibfnamefont
  {A.}~\bibnamefont {Miyake}}, \bibinfo {author} {\bibfnamefont
  {K.}~\bibnamefont {Akiba}}, \bibinfo {author} {\bibfnamefont
  {M.}~\bibnamefont {Tokunaga}}, \bibinfo {author} {\bibfnamefont
  {A.}~\bibnamefont {Matsuo}}, \bibinfo {author} {\bibfnamefont
  {K.}~\bibnamefont {Kindo}}, \bibinfo {author} {\bibfnamefont
  {T.}~\bibnamefont {Koretsune}}, \bibinfo {author} {\bibfnamefont
  {R.}~\bibnamefont {Arita}},\ and\ \bibinfo {author} {\bibfnamefont
  {Y.}~\bibnamefont {Tokura}},\ }\bibfield  {title} {\bibinfo {title}
  {Topological transitions among skyrmion- and hedgehog-lattice states in cubic
  chiral magnets},\ }\href {https://doi.org/10.1038/s41467-019-08985-6}
  {\bibfield  {journal} {\bibinfo  {journal} {Nat. Commun.}\ }\textbf {\bibinfo
  {volume} {10}},\ \bibinfo {pages} {1} (\bibinfo {year} {2019})}\BibitemShut
  {NoStop}%
\bibitem [{\citenamefont {G\"obel}\ \emph {et~al.}(2020)\citenamefont
  {G\"obel}, \citenamefont {Akosa}, \citenamefont {Tatara},\ and\ \citenamefont
  {Mertig}}]{gobel_topological_2020}%
  \BibitemOpen
  \bibfield  {author} {\bibinfo {author} {\bibfnamefont {B.}~\bibnamefont
  {G\"obel}}, \bibinfo {author} {\bibfnamefont {C.~A.}\ \bibnamefont {Akosa}},
  \bibinfo {author} {\bibfnamefont {G.}~\bibnamefont {Tatara}},\ and\ \bibinfo
  {author} {\bibfnamefont {I.}~\bibnamefont {Mertig}},\ }\bibfield  {title}
  {\bibinfo {title} {Topological {Hall} signatures of magnetic hopfions},\
  }\href {https://doi.org/10.1103/PhysRevResearch.2.013315} {\bibfield
  {journal} {\bibinfo  {journal} {Phys. Rev. Research}\ }\textbf {\bibinfo
  {volume} {2}},\ \bibinfo {pages} {013315} (\bibinfo {year}
  {2020})}\BibitemShut {NoStop}%
\bibitem [{\citenamefont {Vistoli}\ \emph {et~al.}(2019)\citenamefont
  {Vistoli}, \citenamefont {Wang}, \citenamefont {Sander}, \citenamefont {Zhu},
  \citenamefont {Casals}, \citenamefont {Cichelero}, \citenamefont
  {Barth\'{e}l\'{e}my}, \citenamefont {Fusil}, \citenamefont {Herranz},
  \citenamefont {Valencia}, \citenamefont {Abrudan}, \citenamefont {Weschke},
  \citenamefont {Nakazawa}, \citenamefont {Kohno}, \citenamefont {Santamaria},
  \citenamefont {Wu}, \citenamefont {Garcia},\ and\ \citenamefont
  {Bibes}}]{vistoli_giant_2019}%
  \BibitemOpen
  \bibfield  {author} {\bibinfo {author} {\bibfnamefont {L.}~\bibnamefont
  {Vistoli}}, \bibinfo {author} {\bibfnamefont {W.}~\bibnamefont {Wang}},
  \bibinfo {author} {\bibfnamefont {A.}~\bibnamefont {Sander}}, \bibinfo
  {author} {\bibfnamefont {Q.}~\bibnamefont {Zhu}}, \bibinfo {author}
  {\bibfnamefont {B.}~\bibnamefont {Casals}}, \bibinfo {author} {\bibfnamefont
  {R.}~\bibnamefont {Cichelero}}, \bibinfo {author} {\bibfnamefont
  {A.}~\bibnamefont {Barth\'{e}l\'{e}my}}, \bibinfo {author} {\bibfnamefont
  {S.}~\bibnamefont {Fusil}}, \bibinfo {author} {\bibfnamefont
  {G.}~\bibnamefont {Herranz}}, \bibinfo {author} {\bibfnamefont
  {S.}~\bibnamefont {Valencia}}, \bibinfo {author} {\bibfnamefont
  {R.}~\bibnamefont {Abrudan}}, \bibinfo {author} {\bibfnamefont
  {E.}~\bibnamefont {Weschke}}, \bibinfo {author} {\bibfnamefont
  {K.}~\bibnamefont {Nakazawa}}, \bibinfo {author} {\bibfnamefont
  {H.}~\bibnamefont {Kohno}}, \bibinfo {author} {\bibfnamefont
  {J.}~\bibnamefont {Santamaria}}, \bibinfo {author} {\bibfnamefont
  {W.}~\bibnamefont {Wu}}, \bibinfo {author} {\bibfnamefont {V.}~\bibnamefont
  {Garcia}},\ and\ \bibinfo {author} {\bibfnamefont {M.}~\bibnamefont
  {Bibes}},\ }\bibfield  {title} {\bibinfo {title} {Giant topological {Hall}
  effect in correlated oxide thin films},\ }\href
  {https://doi.org/10.1038/s41567-018-0307-5} {\bibfield  {journal} {\bibinfo
  {journal} {Nat. Phys.}\ }\textbf {\bibinfo {volume} {15}},\ \bibinfo {pages}
  {67} (\bibinfo {year} {2019})}\BibitemShut {NoStop}%
\bibitem [{\citenamefont {Kneidinger}\ \emph {et~al.}(2014)\citenamefont
  {Kneidinger}, \citenamefont {Salamakha}, \citenamefont {Bauer}, \citenamefont
  {Zeiringer}, \citenamefont {Rogl}, \citenamefont {Blaas-Schenner},
  \citenamefont {Reith},\ and\ \citenamefont {Podloucky}}]{Kneidinger2014}%
  \BibitemOpen
  \bibfield  {author} {\bibinfo {author} {\bibfnamefont {F.}~\bibnamefont
  {Kneidinger}}, \bibinfo {author} {\bibfnamefont {L.}~\bibnamefont
  {Salamakha}}, \bibinfo {author} {\bibfnamefont {E.}~\bibnamefont {Bauer}},
  \bibinfo {author} {\bibfnamefont {I.}~\bibnamefont {Zeiringer}}, \bibinfo
  {author} {\bibfnamefont {P.}~\bibnamefont {Rogl}}, \bibinfo {author}
  {\bibfnamefont {C.}~\bibnamefont {Blaas-Schenner}}, \bibinfo {author}
  {\bibfnamefont {D.}~\bibnamefont {Reith}},\ and\ \bibinfo {author}
  {\bibfnamefont {R.}~\bibnamefont {Podloucky}},\ }\bibfield  {title} {\bibinfo
  {title} {Superconductivity in noncentrosymmetric {Ba}{Al}$_4$ derived
  structures},\ }\href {https://doi.org/10.1103/PhysRevB.90.024504} {\bibfield
  {journal} {\bibinfo  {journal} {Phys. Rev. B}\ }\textbf {\bibinfo {volume}
  {90}},\ \bibinfo {pages} {024504} (\bibinfo {year} {2014})}\BibitemShut
  {NoStop}%
\bibitem [{\citenamefont {Wang}\ \emph {et~al.}(2020)\citenamefont {Wang},
  \citenamefont {Mori}, \citenamefont {Wang}, \citenamefont {Wang},
  \citenamefont {Ma}, \citenamefont {Latzke}, \citenamefont {Graf},
  \citenamefont {Denlinger}, \citenamefont {Campbell}, \citenamefont
  {Bernevig}, \citenamefont {Lanzara},\ and\ \citenamefont
  {Paglione}}]{wang_crystalline_2020}%
  \BibitemOpen
  \bibfield  {author} {\bibinfo {author} {\bibfnamefont {K.}~\bibnamefont
  {Wang}}, \bibinfo {author} {\bibfnamefont {R.}~\bibnamefont {Mori}}, \bibinfo
  {author} {\bibfnamefont {Z.}~\bibnamefont {Wang}}, \bibinfo {author}
  {\bibfnamefont {L.}~\bibnamefont {Wang}}, \bibinfo {author} {\bibfnamefont
  {J.~H.~S.}\ \bibnamefont {Ma}}, \bibinfo {author} {\bibfnamefont {D.~W.}\
  \bibnamefont {Latzke}}, \bibinfo {author} {\bibfnamefont {D.~E.}\
  \bibnamefont {Graf}}, \bibinfo {author} {\bibfnamefont {J.~D.}\ \bibnamefont
  {Denlinger}}, \bibinfo {author} {\bibfnamefont {D.}~\bibnamefont {Campbell}},
  \bibinfo {author} {\bibfnamefont {B.~A.}\ \bibnamefont {Bernevig}}, \bibinfo
  {author} {\bibfnamefont {A.}~\bibnamefont {Lanzara}},\ and\ \bibinfo {author}
  {\bibfnamefont {J.}~\bibnamefont {Paglione}},\ }\bibfield  {title} {\bibinfo
  {title} {Crystalline symmetry-protected non-trivial topology in prototype
  compound {BaAl$_4$}},\ }\href {http://arxiv.org/abs/2007.12571} {\bibfield
  {journal} {\bibinfo  {journal} {arXiv:2007.12571}\ } (\bibinfo {year}
  {2020})}\BibitemShut {NoStop}%
\bibitem [{\citenamefont {Nakamura}\ \emph {et~al.}(2016)\citenamefont
  {Nakamura}, \citenamefont {Uejo}, \citenamefont {Harima}, \citenamefont
  {Araki}, \citenamefont {Kobayashi}, \citenamefont {Nakashima}, \citenamefont
  {Amako}, \citenamefont {Hedo}, \citenamefont {Nakama},\ and\ \citenamefont
  {Ōnuki}}]{Nakamura2016}%
  \BibitemOpen
  \bibfield  {author} {\bibinfo {author} {\bibfnamefont {A.}~\bibnamefont
  {Nakamura}}, \bibinfo {author} {\bibfnamefont {T.}~\bibnamefont {Uejo}},
  \bibinfo {author} {\bibfnamefont {H.}~\bibnamefont {Harima}}, \bibinfo
  {author} {\bibfnamefont {S.}~\bibnamefont {Araki}}, \bibinfo {author}
  {\bibfnamefont {T.~C.}\ \bibnamefont {Kobayashi}}, \bibinfo {author}
  {\bibfnamefont {M.}~\bibnamefont {Nakashima}}, \bibinfo {author}
  {\bibfnamefont {Y.}~\bibnamefont {Amako}}, \bibinfo {author} {\bibfnamefont
  {M.}~\bibnamefont {Hedo}}, \bibinfo {author} {\bibfnamefont {T.}~\bibnamefont
  {Nakama}},\ and\ \bibinfo {author} {\bibfnamefont {Y.}~\bibnamefont
  {Ōnuki}},\ }\bibfield  {title} {\bibinfo {title} {Characteristic {Fermi}
  surfaces and charge density wave in {Sr}{Al}$_4$ and related compounds with
  the {Ba}{Al}$_4$-type tetragonal structure},\ }\href
  {https://doi.org/10.1016/j.jallcom.2015.08.193} {\bibfield  {journal}
  {\bibinfo  {journal} {J. Alloys Compd.}\ }\textbf {\bibinfo {volume} {654}},\
  \bibinfo {pages} {290} (\bibinfo {year} {2016})}\BibitemShut {NoStop}%
\bibitem [{\citenamefont {Nakamura}\ \emph {et~al.}(2013)\citenamefont
  {Nakamura}, \citenamefont {Hiranaka}, \citenamefont {Hedo}, \citenamefont
  {Nakama}, \citenamefont {Miura}, \citenamefont {Tsutsumi}, \citenamefont
  {Mori}, \citenamefont {Ishida}, \citenamefont {Mitamura}, \citenamefont
  {Hirose}, \citenamefont {Sugiyama}, \citenamefont {Honda}, \citenamefont
  {Settai}, \citenamefont {Takeuchi}, \citenamefont {Hagiwara}, \citenamefont
  {D.~Matsuda}, \citenamefont {Yamamoto}, \citenamefont {Haga}, \citenamefont
  {Matsubayashi}, \citenamefont {Uwatoko}, \citenamefont {Harima},\ and\
  \citenamefont {\={O}nuki}}]{nakamura_magnetic_2013}%
  \BibitemOpen
  \bibfield  {author} {\bibinfo {author} {\bibfnamefont {A.}~\bibnamefont
  {Nakamura}}, \bibinfo {author} {\bibfnamefont {Y.}~\bibnamefont {Hiranaka}},
  \bibinfo {author} {\bibfnamefont {M.}~\bibnamefont {Hedo}}, \bibinfo {author}
  {\bibfnamefont {T.}~\bibnamefont {Nakama}}, \bibinfo {author} {\bibfnamefont
  {Y.}~\bibnamefont {Miura}}, \bibinfo {author} {\bibfnamefont
  {H.}~\bibnamefont {Tsutsumi}}, \bibinfo {author} {\bibfnamefont
  {A.}~\bibnamefont {Mori}}, \bibinfo {author} {\bibfnamefont {K.}~\bibnamefont
  {Ishida}}, \bibinfo {author} {\bibfnamefont {K.}~\bibnamefont {Mitamura}},
  \bibinfo {author} {\bibfnamefont {Y.}~\bibnamefont {Hirose}}, \bibinfo
  {author} {\bibfnamefont {K.}~\bibnamefont {Sugiyama}}, \bibinfo {author}
  {\bibfnamefont {F.}~\bibnamefont {Honda}}, \bibinfo {author} {\bibfnamefont
  {R.}~\bibnamefont {Settai}}, \bibinfo {author} {\bibfnamefont
  {T.}~\bibnamefont {Takeuchi}}, \bibinfo {author} {\bibfnamefont
  {M.}~\bibnamefont {Hagiwara}}, \bibinfo {author} {\bibfnamefont
  {T.}~\bibnamefont {D.~Matsuda}}, \bibinfo {author} {\bibfnamefont
  {E.}~\bibnamefont {Yamamoto}}, \bibinfo {author} {\bibfnamefont
  {Y.}~\bibnamefont {Haga}}, \bibinfo {author} {\bibfnamefont {K.}~\bibnamefont
  {Matsubayashi}}, \bibinfo {author} {\bibfnamefont {Y.}~\bibnamefont
  {Uwatoko}}, \bibinfo {author} {\bibfnamefont {H.}~\bibnamefont {Harima}},\
  and\ \bibinfo {author} {\bibfnamefont {Y.}~\bibnamefont {\={O}nuki}},\
  }\bibfield  {title} {\bibinfo {title} {Magnetic and {F}ermi surface
  properties of {EuGa}$_4$},\ }\href {https://doi.org/10.7566/JPSJ.82.104703}
  {\bibfield  {journal} {\bibinfo  {journal} {J. Phys. Soc. Jpn.}\ }\textbf
  {\bibinfo {volume} {82}},\ \bibinfo {pages} {104703} (\bibinfo {year}
  {2013})}\BibitemShut {NoStop}%
\bibitem [{\citenamefont {Araki}\ \emph {et~al.}(2013)\citenamefont {Araki},
  \citenamefont {Ikeda}, \citenamefont {Kobayashi}, \citenamefont {Nakamura},
  \citenamefont {Hiranaka}, \citenamefont {Hedo}, \citenamefont {Nakama},\ and\
  \citenamefont {\={O}nuki}}]{araki_charge_2013}%
  \BibitemOpen
  \bibfield  {author} {\bibinfo {author} {\bibfnamefont {S.}~\bibnamefont
  {Araki}}, \bibinfo {author} {\bibfnamefont {Y.}~\bibnamefont {Ikeda}},
  \bibinfo {author} {\bibfnamefont {T.~C.}\ \bibnamefont {Kobayashi}}, \bibinfo
  {author} {\bibfnamefont {A.}~\bibnamefont {Nakamura}}, \bibinfo {author}
  {\bibfnamefont {Y.}~\bibnamefont {Hiranaka}}, \bibinfo {author}
  {\bibfnamefont {M.}~\bibnamefont {Hedo}}, \bibinfo {author} {\bibfnamefont
  {T.}~\bibnamefont {Nakama}},\ and\ \bibinfo {author} {\bibfnamefont
  {Y.}~\bibnamefont {\={O}nuki}},\ }\bibfield  {title} {\bibinfo {title}
  {Charge density wave transition in {EuAl$_4$}},\ }\href
  {https://doi.org/10.7566/JPSJ.83.015001} {\bibfield  {journal} {\bibinfo
  {journal} {J. Phys. Soc. Jpn.}\ }\textbf {\bibinfo {volume} {83}},\ \bibinfo
  {pages} {015001} (\bibinfo {year} {2013})}\BibitemShut {NoStop}%
\bibitem [{\citenamefont {Nakamura}\ \emph {et~al.}(2014)\citenamefont
  {Nakamura}, \citenamefont {Hiranaka}, \citenamefont {Hedo}, \citenamefont
  {Nakama}, \citenamefont {Miura}, \citenamefont {Tsutsumi}, \citenamefont
  {Mori}, \citenamefont {Ishida}, \citenamefont {Mitamura}, \citenamefont
  {Hirose}, \citenamefont {Sugiyama}, \citenamefont {Honda}, \citenamefont
  {Takeuchi}, \citenamefont {Matsuda}, \citenamefont {Yamamoto}, \citenamefont
  {Haga},\ and\ \citenamefont {\={O}nuki}}]{nakamura_unique_2014}%
  \BibitemOpen
  \bibfield  {author} {\bibinfo {author} {\bibfnamefont {A.}~\bibnamefont
  {Nakamura}}, \bibinfo {author} {\bibfnamefont {Y.}~\bibnamefont {Hiranaka}},
  \bibinfo {author} {\bibfnamefont {M.}~\bibnamefont {Hedo}}, \bibinfo {author}
  {\bibfnamefont {T.}~\bibnamefont {Nakama}}, \bibinfo {author} {\bibfnamefont
  {Y.}~\bibnamefont {Miura}}, \bibinfo {author} {\bibfnamefont
  {H.}~\bibnamefont {Tsutsumi}}, \bibinfo {author} {\bibfnamefont
  {A.}~\bibnamefont {Mori}}, \bibinfo {author} {\bibfnamefont {K.}~\bibnamefont
  {Ishida}}, \bibinfo {author} {\bibfnamefont {K.}~\bibnamefont {Mitamura}},
  \bibinfo {author} {\bibfnamefont {Y.}~\bibnamefont {Hirose}}, \bibinfo
  {author} {\bibfnamefont {K.}~\bibnamefont {Sugiyama}}, \bibinfo {author}
  {\bibfnamefont {F.}~\bibnamefont {Honda}}, \bibinfo {author} {\bibfnamefont
  {T.}~\bibnamefont {Takeuchi}}, \bibinfo {author} {\bibfnamefont {T.~D.}\
  \bibnamefont {Matsuda}}, \bibinfo {author} {\bibfnamefont {E.}~\bibnamefont
  {Yamamoto}}, \bibinfo {author} {\bibfnamefont {Y.}~\bibnamefont {Haga}},\
  and\ \bibinfo {author} {\bibfnamefont {Y.}~\bibnamefont {\={O}nuki}},\
  }\bibfield  {title} {\bibinfo {title} {Unique {Fermi} surface and emergence
  of charge density wave in {EuGa$_4$} and {EuAl$_4$}},\ }\href
  {https://doi.org/10.7566/JPSCP.3.011012} {\bibfield  {journal} {\bibinfo
  {journal} {JPS Conf. Proc.}\ }\textbf {\bibinfo {volume} {3}},\ \bibinfo
  {pages} {011012} (\bibinfo {year} {2014})}\BibitemShut {NoStop}%
\bibitem [{\citenamefont {Nakamura}\ \emph {et~al.}(2015)\citenamefont
  {Nakamura}, \citenamefont {Uejo}, \citenamefont {Honda}, \citenamefont
  {Takeuchi}, \citenamefont {Harima}, \citenamefont {Yamamoto}, \citenamefont
  {Haga}, \citenamefont {Matsubayashi}, \citenamefont {Uwatoko}, \citenamefont
  {Hedo}, \citenamefont {Nakama},\ and\ \citenamefont
  {\={O}nuki}}]{nakamura_transport_2015}%
  \BibitemOpen
  \bibfield  {author} {\bibinfo {author} {\bibfnamefont {A.}~\bibnamefont
  {Nakamura}}, \bibinfo {author} {\bibfnamefont {T.}~\bibnamefont {Uejo}},
  \bibinfo {author} {\bibfnamefont {F.}~\bibnamefont {Honda}}, \bibinfo
  {author} {\bibfnamefont {T.}~\bibnamefont {Takeuchi}}, \bibinfo {author}
  {\bibfnamefont {H.}~\bibnamefont {Harima}}, \bibinfo {author} {\bibfnamefont
  {E.}~\bibnamefont {Yamamoto}}, \bibinfo {author} {\bibfnamefont
  {Y.}~\bibnamefont {Haga}}, \bibinfo {author} {\bibfnamefont {K.}~\bibnamefont
  {Matsubayashi}}, \bibinfo {author} {\bibfnamefont {Y.}~\bibnamefont
  {Uwatoko}}, \bibinfo {author} {\bibfnamefont {M.}~\bibnamefont {Hedo}},
  \bibinfo {author} {\bibfnamefont {T.}~\bibnamefont {Nakama}},\ and\ \bibinfo
  {author} {\bibfnamefont {Y.}~\bibnamefont {\={O}nuki}},\ }\bibfield  {title}
  {\bibinfo {title} {Transport and magnetic properties of {EuAl$_4$} and
  {EuGa$_4$}},\ }\href {https://doi.org/10.7566/JPSJ.84.124711} {\bibfield
  {journal} {\bibinfo  {journal} {J. Phys. Soc. Jpn.}\ }\textbf {\bibinfo
  {volume} {84}},\ \bibinfo {pages} {124711} (\bibinfo {year}
  {2015})}\BibitemShut {NoStop}%
\bibitem [{\citenamefont {Shimomura}\ \emph {et~al.}(2019)\citenamefont
  {Shimomura}, \citenamefont {Murao}, \citenamefont {Tsutsui}, \citenamefont
  {Nakao}, \citenamefont {Nakamura}, \citenamefont {Hedo}, \citenamefont
  {Nakama},\ and\ \citenamefont {\={O}nuki}}]{shimomura_lattice_2019}%
  \BibitemOpen
  \bibfield  {author} {\bibinfo {author} {\bibfnamefont {S.}~\bibnamefont
  {Shimomura}}, \bibinfo {author} {\bibfnamefont {H.}~\bibnamefont {Murao}},
  \bibinfo {author} {\bibfnamefont {S.}~\bibnamefont {Tsutsui}}, \bibinfo
  {author} {\bibfnamefont {H.}~\bibnamefont {Nakao}}, \bibinfo {author}
  {\bibfnamefont {A.}~\bibnamefont {Nakamura}}, \bibinfo {author}
  {\bibfnamefont {M.}~\bibnamefont {Hedo}}, \bibinfo {author} {\bibfnamefont
  {T.}~\bibnamefont {Nakama}},\ and\ \bibinfo {author} {\bibfnamefont
  {Y.}~\bibnamefont {\={O}nuki}},\ }\bibfield  {title} {\bibinfo {title}
  {Lattice modulation and structural phase transition in the antiferromagnet
  {EuAl$_4$}},\ }\href {https://doi.org/10.7566/JPSJ.88.014602} {\bibfield
  {journal} {\bibinfo  {journal} {J. Phys. Soc. Jpn.}\ }\textbf {\bibinfo
  {volume} {88}},\ \bibinfo {pages} {014602} (\bibinfo {year}
  {2019})}\BibitemShut {NoStop}%
\bibitem [{\citenamefont {Kobata}\ \emph {et~al.}(2016)\citenamefont {Kobata},
  \citenamefont {Fujimori}, \citenamefont {Takeda}, \citenamefont {Okane},
  \citenamefont {Saitoh}, \citenamefont {Kobayashi}, \citenamefont {Yamagami},
  \citenamefont {Nakamura}, \citenamefont {Hedo}, \citenamefont {Nakama},\ and\
  \citenamefont {\={O}nuki}}]{Kobata2016}%
  \BibitemOpen
  \bibfield  {author} {\bibinfo {author} {\bibfnamefont {M.}~\bibnamefont
  {Kobata}}, \bibinfo {author} {\bibfnamefont {S.}~\bibnamefont {Fujimori}},
  \bibinfo {author} {\bibfnamefont {Y.}~\bibnamefont {Takeda}}, \bibinfo
  {author} {\bibfnamefont {T.}~\bibnamefont {Okane}}, \bibinfo {author}
  {\bibfnamefont {Y.}~\bibnamefont {Saitoh}}, \bibinfo {author} {\bibfnamefont
  {K.}~\bibnamefont {Kobayashi}}, \bibinfo {author} {\bibfnamefont
  {H.}~\bibnamefont {Yamagami}}, \bibinfo {author} {\bibfnamefont
  {A.}~\bibnamefont {Nakamura}}, \bibinfo {author} {\bibfnamefont
  {M.}~\bibnamefont {Hedo}}, \bibinfo {author} {\bibfnamefont {T.}~\bibnamefont
  {Nakama}},\ and\ \bibinfo {author} {\bibfnamefont {Y.}~\bibnamefont
  {\={O}nuki}},\ }\bibfield  {title} {\bibinfo {title} {Electronic structure of
  {Eu}{Al}$_4$ studied by photoelectron spectroscopy},\ }\href
  {https://doi.org/10.7566/JPSJ.85.094703} {\bibfield  {journal} {\bibinfo
  {journal} {J. Phys. Soc. Jpn.}\ }\textbf {\bibinfo {volume} {85}},\ \bibinfo
  {pages} {094703} (\bibinfo {year} {2016})}\BibitemShut {NoStop}%
\bibitem [{\citenamefont {Fawcett}\ and\ \citenamefont
  {Reed}(1963)}]{Fawcett_1963}%
  \BibitemOpen
  \bibfield  {author} {\bibinfo {author} {\bibfnamefont {E.}~\bibnamefont
  {Fawcett}}\ and\ \bibinfo {author} {\bibfnamefont {W.~A.}\ \bibnamefont
  {Reed}},\ }\bibfield  {title} {\bibinfo {title} {Effects of compensation on
  the galvanomagnetic properties of nonmagnetic and ferromagnetic metals},\
  }\href {https://doi.org/10.1103/PhysRev.131.2463} {\bibfield  {journal}
  {\bibinfo  {journal} {Phys. Rev.}\ }\textbf {\bibinfo {volume} {131}},\
  \bibinfo {pages} {2463} (\bibinfo {year} {1963})}\BibitemShut {NoStop}%
\bibitem [{\citenamefont {Nagaosa}\ \emph {et~al.}(2010)\citenamefont
  {Nagaosa}, \citenamefont {Sinova}, \citenamefont {Onoda}, \citenamefont
  {MacDonald},\ and\ \citenamefont {Ong}}]{nagaosa_anomalous_2010}%
  \BibitemOpen
  \bibfield  {author} {\bibinfo {author} {\bibfnamefont {N.}~\bibnamefont
  {Nagaosa}}, \bibinfo {author} {\bibfnamefont {J.}~\bibnamefont {Sinova}},
  \bibinfo {author} {\bibfnamefont {S.}~\bibnamefont {Onoda}}, \bibinfo
  {author} {\bibfnamefont {A.~H.}\ \bibnamefont {MacDonald}},\ and\ \bibinfo
  {author} {\bibfnamefont {N.~P.}\ \bibnamefont {Ong}},\ }\bibfield  {title}
  {\bibinfo {title} {Anomalous {Hall} effect},\ }\href
  {https://doi.org/10.1103/RevModPhys.82.1539} {\bibfield  {journal} {\bibinfo
  {journal} {Rev. Mod. Phys.}\ }\textbf {\bibinfo {volume} {82}},\ \bibinfo
  {pages} {1539} (\bibinfo {year} {2010})}\BibitemShut {NoStop}%
\bibitem [{\citenamefont {Lee}\ \emph {et~al.}(2007)\citenamefont {Lee},
  \citenamefont {Onose}, \citenamefont {Tokura},\ and\ \citenamefont
  {Ong}}]{lee_hidden_2007}%
  \BibitemOpen
  \bibfield  {author} {\bibinfo {author} {\bibfnamefont {M.}~\bibnamefont
  {Lee}}, \bibinfo {author} {\bibfnamefont {Y.}~\bibnamefont {Onose}}, \bibinfo
  {author} {\bibfnamefont {Y.}~\bibnamefont {Tokura}},\ and\ \bibinfo {author}
  {\bibfnamefont {N.~P.}\ \bibnamefont {Ong}},\ }\bibfield  {title} {\bibinfo
  {title} {Hidden constant in the anomalous {Hall} effect of high-purity magnet
  {MnSi}},\ }\href {https://doi.org/10.1103/PhysRevB.75.172403} {\bibfield
  {journal} {\bibinfo  {journal} {Phys. Rev. B}\ }\textbf {\bibinfo {volume}
  {75}},\ \bibinfo {pages} {172403} (\bibinfo {year} {2007})}\BibitemShut
  {NoStop}%
\bibitem [{\citenamefont {Tian}\ \emph {et~al.}(2009)\citenamefont {Tian},
  \citenamefont {Ye},\ and\ \citenamefont {Jin}}]{tian_proper_2009}%
  \BibitemOpen
  \bibfield  {author} {\bibinfo {author} {\bibfnamefont {Y.}~\bibnamefont
  {Tian}}, \bibinfo {author} {\bibfnamefont {L.}~\bibnamefont {Ye}},\ and\
  \bibinfo {author} {\bibfnamefont {X.}~\bibnamefont {Jin}},\ }\bibfield
  {title} {\bibinfo {title} {Proper scaling of the anomalous {Hall} effect},\
  }\href {https://doi.org/10.1103/PhysRevLett.103.087206} {\bibfield  {journal}
  {\bibinfo  {journal} {Phys. Rev. Lett.}\ }\textbf {\bibinfo {volume} {103}},\
  \bibinfo {pages} {087206} (\bibinfo {year} {2009})}\BibitemShut {NoStop}%
\bibitem [{\citenamefont {Hou}\ \emph {et~al.}(2015)\citenamefont {Hou},
  \citenamefont {Su}, \citenamefont {Tian}, \citenamefont {Jin}, \citenamefont
  {Yang},\ and\ \citenamefont {Niu}}]{hou_multivariable_2015}%
  \BibitemOpen
  \bibfield  {author} {\bibinfo {author} {\bibfnamefont {D.}~\bibnamefont
  {Hou}}, \bibinfo {author} {\bibfnamefont {G.}~\bibnamefont {Su}}, \bibinfo
  {author} {\bibfnamefont {Y.}~\bibnamefont {Tian}}, \bibinfo {author}
  {\bibfnamefont {X.}~\bibnamefont {Jin}}, \bibinfo {author} {\bibfnamefont
  {S.~A.}\ \bibnamefont {Yang}},\ and\ \bibinfo {author} {\bibfnamefont
  {Q.}~\bibnamefont {Niu}},\ }\bibfield  {title} {\bibinfo {title}
  {Multivariable scaling for the anomalous {Hall} effect},\ }\href
  {https://doi.org/10.1103/PhysRevLett.114.217203} {\bibfield  {journal}
  {\bibinfo  {journal} {Phys. Rev. Lett.}\ }\textbf {\bibinfo {volume} {114}},\
  \bibinfo {pages} {217203} (\bibinfo {year} {2015})}\BibitemShut {NoStop}%
\bibitem [{\citenamefont {Ying}\ \emph {et~al.}(2012)\citenamefont {Ying},
  \citenamefont {Yan}, \citenamefont {Wang}, \citenamefont {Xiang},
  \citenamefont {Cheng}, \citenamefont {Ye},\ and\ \citenamefont
  {Chen}}]{Ying2012}%
  \BibitemOpen
  \bibfield  {author} {\bibinfo {author} {\bibfnamefont {J.~J.}\ \bibnamefont
  {Ying}}, \bibinfo {author} {\bibfnamefont {Y.~J.}\ \bibnamefont {Yan}},
  \bibinfo {author} {\bibfnamefont {A.~F.}\ \bibnamefont {Wang}}, \bibinfo
  {author} {\bibfnamefont {Z.~J.}\ \bibnamefont {Xiang}}, \bibinfo {author}
  {\bibfnamefont {P.}~\bibnamefont {Cheng}}, \bibinfo {author} {\bibfnamefont
  {G.~J.}\ \bibnamefont {Ye}},\ and\ \bibinfo {author} {\bibfnamefont {X.~H.}\
  \bibnamefont {Chen}},\ }\bibfield  {title} {\bibinfo {title} {Metamagnetic
  transition in {Ca}$_{1-x}${Sr}$_{x}${Co}$_2${As}$_2$ ($x$ = 0 and 0.1) single
  crystals},\ }\href {https://doi.org/10.1103/PhysRevB.85.214414} {\bibfield
  {journal} {\bibinfo  {journal} {Phys. Rev. B}\ }\textbf {\bibinfo {volume}
  {85}},\ \bibinfo {pages} {214414} (\bibinfo {year} {2012})}\BibitemShut
  {NoStop}%
\bibitem [{\citenamefont {Kakihana}\ \emph {et~al.}(2017)\citenamefont
  {Kakihana}, \citenamefont {Aoki}, \citenamefont {Nakamura}, \citenamefont
  {Honda}, \citenamefont {Nakashima}, \citenamefont {Amako}, \citenamefont
  {Nakamura}, \citenamefont {Sakakibara}, \citenamefont {Hedo}, \citenamefont
  {Nakama},\ and\ \citenamefont {\={O}nuki}}]{kakihana_giant_2017}%
  \BibitemOpen
  \bibfield  {author} {\bibinfo {author} {\bibfnamefont {M.}~\bibnamefont
  {Kakihana}}, \bibinfo {author} {\bibfnamefont {D.}~\bibnamefont {Aoki}},
  \bibinfo {author} {\bibfnamefont {A.}~\bibnamefont {Nakamura}}, \bibinfo
  {author} {\bibfnamefont {F.}~\bibnamefont {Honda}}, \bibinfo {author}
  {\bibfnamefont {M.}~\bibnamefont {Nakashima}}, \bibinfo {author}
  {\bibfnamefont {Y.}~\bibnamefont {Amako}}, \bibinfo {author} {\bibfnamefont
  {S.}~\bibnamefont {Nakamura}}, \bibinfo {author} {\bibfnamefont
  {T.}~\bibnamefont {Sakakibara}}, \bibinfo {author} {\bibfnamefont
  {M.}~\bibnamefont {Hedo}}, \bibinfo {author} {\bibfnamefont {T.}~\bibnamefont
  {Nakama}},\ and\ \bibinfo {author} {\bibfnamefont {Y.}~\bibnamefont
  {\={O}nuki}},\ }\bibfield  {title} {\bibinfo {title} {Giant {Hall}
  resistivity and magnetoresistance in cubic chiral antiferromagnet {EuPtSi}},\
  }\href {https://doi.org/10.7566/JPSJ.87.023701} {\bibfield  {journal}
  {\bibinfo  {journal} {J. Phys. Soc. Jpn.}\ }\textbf {\bibinfo {volume}
  {87}},\ \bibinfo {pages} {023701} (\bibinfo {year} {2017})}\BibitemShut
  {NoStop}%
\bibitem [{\citenamefont {Xu}\ \emph {et~al.}(2020)\citenamefont {Xu},
  \citenamefont {Das}, \citenamefont {Ma}, \citenamefont {Yi}, \citenamefont
  {Shi}, \citenamefont {Tiwari}, \citenamefont {Tsirkin}, \citenamefont
  {Neupert}, \citenamefont {Medarde}, \citenamefont {Shi}, \citenamefont
  {Chang},\ and\ \citenamefont {Shang}}]{xu_topological_2020}%
  \BibitemOpen
  \bibfield  {author} {\bibinfo {author} {\bibfnamefont {Y.}~\bibnamefont
  {Xu}}, \bibinfo {author} {\bibfnamefont {L.}~\bibnamefont {Das}}, \bibinfo
  {author} {\bibfnamefont {J.~Z.}\ \bibnamefont {Ma}}, \bibinfo {author}
  {\bibfnamefont {C.~J.}\ \bibnamefont {Yi}}, \bibinfo {author} {\bibfnamefont
  {Y.~G.}\ \bibnamefont {Shi}}, \bibinfo {author} {\bibfnamefont
  {A.}~\bibnamefont {Tiwari}}, \bibinfo {author} {\bibfnamefont {S.~S.}\
  \bibnamefont {Tsirkin}}, \bibinfo {author} {\bibfnamefont {T.}~\bibnamefont
  {Neupert}}, \bibinfo {author} {\bibfnamefont {M.}~\bibnamefont {Medarde}},
  \bibinfo {author} {\bibfnamefont {M.}~\bibnamefont {Shi}}, \bibinfo {author}
  {\bibfnamefont {J.}~\bibnamefont {Chang}},\ and\ \bibinfo {author}
  {\bibfnamefont {T.}~\bibnamefont {Shang}},\ }\bibfield  {title} {\bibinfo
  {title} {Topological transverse transport in the presence and absence of
  long-range magnetic order in {EuCd$_2$As$_2$}},\ }\href
  {http://arxiv.org/abs/2008.05390} {\bibfield  {journal} {\bibinfo  {journal}
  {arXiv:2008.05390}\ } (\bibinfo {year} {2020})}\BibitemShut {NoStop}%
\bibitem [{\citenamefont {Puphal}\ \emph {et~al.}(2020)\citenamefont {Puphal},
  \citenamefont {Pomjakushin}, \citenamefont {Kanazawa}, \citenamefont
  {Ukleev}, \citenamefont {Gawryluk}, \citenamefont {Ma}, \citenamefont
  {Naamneh}, \citenamefont {Plumb}, \citenamefont {Keller}, \citenamefont
  {Cubitt}, \citenamefont {Pomjakushina},\ and\ \citenamefont
  {White}}]{puphal_topological_2020}%
  \BibitemOpen
  \bibfield  {author} {\bibinfo {author} {\bibfnamefont {P.}~\bibnamefont
  {Puphal}}, \bibinfo {author} {\bibfnamefont {V.}~\bibnamefont {Pomjakushin}},
  \bibinfo {author} {\bibfnamefont {N.}~\bibnamefont {Kanazawa}}, \bibinfo
  {author} {\bibfnamefont {V.}~\bibnamefont {Ukleev}}, \bibinfo {author}
  {\bibfnamefont {D.~J.}\ \bibnamefont {Gawryluk}}, \bibinfo {author}
  {\bibfnamefont {J.}~\bibnamefont {Ma}}, \bibinfo {author} {\bibfnamefont
  {M.}~\bibnamefont {Naamneh}}, \bibinfo {author} {\bibfnamefont {N.~C.}\
  \bibnamefont {Plumb}}, \bibinfo {author} {\bibfnamefont {L.}~\bibnamefont
  {Keller}}, \bibinfo {author} {\bibfnamefont {R.}~\bibnamefont {Cubitt}},
  \bibinfo {author} {\bibfnamefont {E.}~\bibnamefont {Pomjakushina}},\ and\
  \bibinfo {author} {\bibfnamefont {J.~S.}\ \bibnamefont {White}},\ }\bibfield
  {title} {\bibinfo {title} {Topological magnetic phase in the candidate {Weyl}
  semimetal {CeAlGe}},\ }\href {https://doi.org/10.1103/PhysRevLett.124.017202}
  {\bibfield  {journal} {\bibinfo  {journal} {Phys. Rev. Lett.}\ }\textbf
  {\bibinfo {volume} {124}},\ \bibinfo {pages} {017202} (\bibinfo {year}
  {2020})}\BibitemShut {NoStop}%
\bibitem [{\citenamefont {M\"uhlbauer}\ \emph {et~al.}(2009)\citenamefont
  {M\"uhlbauer}, \citenamefont {Binz}, \citenamefont {Jonietz}, \citenamefont
  {Pfleiderer}, \citenamefont {Rosch}, \citenamefont {Neubauer}, \citenamefont
  {Georgii},\ and\ \citenamefont {B\"oni}}]{muhlbauer_skyrmion_2009}%
  \BibitemOpen
  \bibfield  {author} {\bibinfo {author} {\bibfnamefont {S.}~\bibnamefont
  {M\"uhlbauer}}, \bibinfo {author} {\bibfnamefont {B.}~\bibnamefont {Binz}},
  \bibinfo {author} {\bibfnamefont {F.}~\bibnamefont {Jonietz}}, \bibinfo
  {author} {\bibfnamefont {C.}~\bibnamefont {Pfleiderer}}, \bibinfo {author}
  {\bibfnamefont {A.}~\bibnamefont {Rosch}}, \bibinfo {author} {\bibfnamefont
  {A.}~\bibnamefont {Neubauer}}, \bibinfo {author} {\bibfnamefont
  {R.}~\bibnamefont {Georgii}},\ and\ \bibinfo {author} {\bibfnamefont
  {P.}~\bibnamefont {B\"oni}},\ }\bibfield  {title} {\bibinfo {title} {Skyrmion
  lattice in a chiral magnet},\ }\href
  {https://doi.org/10.1126/science.1166767} {\bibfield  {journal} {\bibinfo
  {journal} {Science}\ }\textbf {\bibinfo {volume} {323}},\ \bibinfo {pages}
  {915} (\bibinfo {year} {2009})}\BibitemShut {NoStop}%
\bibitem [{\citenamefont {Yu}\ \emph {et~al.}(2011)\citenamefont {Yu},
  \citenamefont {Kanazawa}, \citenamefont {Onose}, \citenamefont {Kimoto},
  \citenamefont {Zhang}, \citenamefont {Ishiwata}, \citenamefont {Matsui},\
  and\ \citenamefont {Tokura}}]{yu_near_2011}%
  \BibitemOpen
  \bibfield  {author} {\bibinfo {author} {\bibfnamefont {X.~Z.}\ \bibnamefont
  {Yu}}, \bibinfo {author} {\bibfnamefont {N.}~\bibnamefont {Kanazawa}},
  \bibinfo {author} {\bibfnamefont {Y.}~\bibnamefont {Onose}}, \bibinfo
  {author} {\bibfnamefont {K.}~\bibnamefont {Kimoto}}, \bibinfo {author}
  {\bibfnamefont {W.~Z.}\ \bibnamefont {Zhang}}, \bibinfo {author}
  {\bibfnamefont {S.}~\bibnamefont {Ishiwata}}, \bibinfo {author}
  {\bibfnamefont {Y.}~\bibnamefont {Matsui}},\ and\ \bibinfo {author}
  {\bibfnamefont {Y.}~\bibnamefont {Tokura}},\ }\bibfield  {title} {\bibinfo
  {title} {Near room-temperature formation of a skyrmion crystal in thin-films
  of the helimagnet {FeGe}},\ }\href {https://doi.org/10.1038/nmat2916}
  {\bibfield  {journal} {\bibinfo  {journal} {Nat. Mater.}\ }\textbf {\bibinfo
  {volume} {10}},\ \bibinfo {pages} {106} (\bibinfo {year} {2011})}\BibitemShut
  {NoStop}%
\bibitem [{\citenamefont {Yu}\ \emph {et~al.}(2010)\citenamefont {Yu},
  \citenamefont {Onose}, \citenamefont {Kanazawa}, \citenamefont {Park},
  \citenamefont {Han}, \citenamefont {Matsui}, \citenamefont {Nagaosa},\ and\
  \citenamefont {Tokura}}]{yu_real-space_2010}%
  \BibitemOpen
  \bibfield  {author} {\bibinfo {author} {\bibfnamefont {X.~Z.}\ \bibnamefont
  {Yu}}, \bibinfo {author} {\bibfnamefont {Y.}~\bibnamefont {Onose}}, \bibinfo
  {author} {\bibfnamefont {N.}~\bibnamefont {Kanazawa}}, \bibinfo {author}
  {\bibfnamefont {J.~H.}\ \bibnamefont {Park}}, \bibinfo {author}
  {\bibfnamefont {J.~H.}\ \bibnamefont {Han}}, \bibinfo {author} {\bibfnamefont
  {Y.}~\bibnamefont {Matsui}}, \bibinfo {author} {\bibfnamefont
  {N.}~\bibnamefont {Nagaosa}},\ and\ \bibinfo {author} {\bibfnamefont
  {Y.}~\bibnamefont {Tokura}},\ }\bibfield  {title} {\bibinfo {title}
  {Real-space observation of a two-dimensional skyrmion crystal},\ }\href
  {https://doi.org/10.1038/nature09124} {\bibfield  {journal} {\bibinfo
  {journal} {Nature}\ }\textbf {\bibinfo {volume} {465}},\ \bibinfo {pages}
  {901} (\bibinfo {year} {2010})}\BibitemShut {NoStop}%
\bibitem [{\citenamefont {Seki}\ \emph
  {et~al.}(2012{\natexlab{a}})\citenamefont {Seki}, \citenamefont {Yu},
  \citenamefont {Ishiwata},\ and\ \citenamefont
  {Tokura}}]{seki_observation_2012}%
  \BibitemOpen
  \bibfield  {author} {\bibinfo {author} {\bibfnamefont {S.}~\bibnamefont
  {Seki}}, \bibinfo {author} {\bibfnamefont {X.~Z.}\ \bibnamefont {Yu}},
  \bibinfo {author} {\bibfnamefont {S.}~\bibnamefont {Ishiwata}},\ and\
  \bibinfo {author} {\bibfnamefont {Y.}~\bibnamefont {Tokura}},\ }\bibfield
  {title} {\bibinfo {title} {Observation of skyrmions in a multiferroic
  material},\ }\href {https://doi.org/10.1126/science.1214143} {\bibfield
  {journal} {\bibinfo  {journal} {Science}\ }\textbf {\bibinfo {volume}
  {336}},\ \bibinfo {pages} {198} (\bibinfo {year}
  {2012}{\natexlab{a}})}\BibitemShut {NoStop}%
\bibitem [{\citenamefont {K\'ezsm\'arki}\ \emph {et~al.}(2015)\citenamefont
  {K\'ezsm\'arki}, \citenamefont {Bord\'acs}, \citenamefont {Milde},
  \citenamefont {Neuber}, \citenamefont {Eng}, \citenamefont {White},
  \citenamefont {R\o{}nnow}, \citenamefont {Dewhurst}, \citenamefont
  {Mochizuki}, \citenamefont {Yanai}, \citenamefont {Nakamura}, \citenamefont
  {Ehlers}, \citenamefont {Tsurkan},\ and\ \citenamefont
  {Loidl}}]{kezsmarki_ne-type_2015}%
  \BibitemOpen
  \bibfield  {author} {\bibinfo {author} {\bibfnamefont {I.}~\bibnamefont
  {K\'ezsm\'arki}}, \bibinfo {author} {\bibfnamefont {S.}~\bibnamefont
  {Bord\'acs}}, \bibinfo {author} {\bibfnamefont {P.}~\bibnamefont {Milde}},
  \bibinfo {author} {\bibfnamefont {E.}~\bibnamefont {Neuber}}, \bibinfo
  {author} {\bibfnamefont {L.~M.}\ \bibnamefont {Eng}}, \bibinfo {author}
  {\bibfnamefont {J.~S.}\ \bibnamefont {White}}, \bibinfo {author}
  {\bibfnamefont {H.~M.}\ \bibnamefont {R\o{}nnow}}, \bibinfo {author}
  {\bibfnamefont {C.~D.}\ \bibnamefont {Dewhurst}}, \bibinfo {author}
  {\bibfnamefont {M.}~\bibnamefont {Mochizuki}}, \bibinfo {author}
  {\bibfnamefont {K.}~\bibnamefont {Yanai}}, \bibinfo {author} {\bibfnamefont
  {H.}~\bibnamefont {Nakamura}}, \bibinfo {author} {\bibfnamefont
  {D.}~\bibnamefont {Ehlers}}, \bibinfo {author} {\bibfnamefont
  {V.}~\bibnamefont {Tsurkan}},\ and\ \bibinfo {author} {\bibfnamefont
  {A.}~\bibnamefont {Loidl}},\ }\bibfield  {title} {\bibinfo {title}
  {N\'eel-type skyrmion lattice with confined orientation in the polar magnetic
  semiconductor {GaV$_4$S$_8$}},\ }\href {https://doi.org/10.1038/nmat4402}
  {\bibfield  {journal} {\bibinfo  {journal} {Nat. Mater.}\ }\textbf {\bibinfo
  {volume} {14}},\ \bibinfo {pages} {1116} (\bibinfo {year}
  {2015})}\BibitemShut {NoStop}%
\bibitem [{\citenamefont {Tokunaga}\ \emph {et~al.}(2015)\citenamefont
  {Tokunaga}, \citenamefont {Yu}, \citenamefont {White}, \citenamefont
  {R\o{}nnow}, \citenamefont {Morikawa}, \citenamefont {Taguchi},\ and\
  \citenamefont {Tokura}}]{tokunaga_new_2015}%
  \BibitemOpen
  \bibfield  {author} {\bibinfo {author} {\bibfnamefont {Y.}~\bibnamefont
  {Tokunaga}}, \bibinfo {author} {\bibfnamefont {X.~Z.}\ \bibnamefont {Yu}},
  \bibinfo {author} {\bibfnamefont {J.~S.}\ \bibnamefont {White}}, \bibinfo
  {author} {\bibfnamefont {H.~M.}\ \bibnamefont {R\o{}nnow}}, \bibinfo {author}
  {\bibfnamefont {D.}~\bibnamefont {Morikawa}}, \bibinfo {author}
  {\bibfnamefont {Y.}~\bibnamefont {Taguchi}},\ and\ \bibinfo {author}
  {\bibfnamefont {Y.}~\bibnamefont {Tokura}},\ }\bibfield  {title} {\bibinfo
  {title} {A new class of chiral materials hosting magnetic skyrmions beyond
  room temperature},\ }\href {https://doi.org/10.1038/ncomms8638} {\bibfield
  {journal} {\bibinfo  {journal} {Nat. Commun.}\ }\textbf {\bibinfo {volume}
  {6}},\ \bibinfo {pages} {1} (\bibinfo {year} {2015})}\BibitemShut {NoStop}%
\bibitem [{\citenamefont {Seki}\ \emph
  {et~al.}(2012{\natexlab{b}})\citenamefont {Seki}, \citenamefont {Kim},
  \citenamefont {Inosov}, \citenamefont {Georgii}, \citenamefont {Keimer},
  \citenamefont {Ishiwata},\ and\ \citenamefont {Tokura}}]{Seki2012}%
  \BibitemOpen
  \bibfield  {author} {\bibinfo {author} {\bibfnamefont {S.}~\bibnamefont
  {Seki}}, \bibinfo {author} {\bibfnamefont {J.-H.}\ \bibnamefont {Kim}},
  \bibinfo {author} {\bibfnamefont {D.~S.}\ \bibnamefont {Inosov}}, \bibinfo
  {author} {\bibfnamefont {R.}~\bibnamefont {Georgii}}, \bibinfo {author}
  {\bibfnamefont {B.}~\bibnamefont {Keimer}}, \bibinfo {author} {\bibfnamefont
  {S.}~\bibnamefont {Ishiwata}},\ and\ \bibinfo {author} {\bibfnamefont
  {Y.}~\bibnamefont {Tokura}},\ }\bibfield  {title} {\bibinfo {title}
  {Formation and rotation of skyrmion crystal in the chiral-lattice insulator
  {Cu$_2$OSeO$_3$}},\ }\href {https://doi.org/10.1103/PhysRevB.85.220406}
  {\bibfield  {journal} {\bibinfo  {journal} {Phys. Rev. B}\ }\textbf {\bibinfo
  {volume} {85}},\ \bibinfo {pages} {220406(R)} (\bibinfo {year}
  {2012}{\natexlab{b}})}\BibitemShut {NoStop}%
\bibitem [{\citenamefont {Hirschberger}\ \emph {et~al.}(2019)\citenamefont
  {Hirschberger}, \citenamefont {Nakajima}, \citenamefont {Gao}, \citenamefont
  {Peng}, \citenamefont {Kikkawa}, \citenamefont {Kurumaji}, \citenamefont
  {Kriener}, \citenamefont {Yamasaki}, \citenamefont {Sagayama}, \citenamefont
  {Nakao}, \citenamefont {Ohishi}, \citenamefont {Kakurai}, \citenamefont
  {Taguchi}, \citenamefont {Yu}, \citenamefont {Arima},\ and\ \citenamefont
  {Tokura}}]{Hirschberger2019}%
  \BibitemOpen
  \bibfield  {author} {\bibinfo {author} {\bibfnamefont {M.}~\bibnamefont
  {Hirschberger}}, \bibinfo {author} {\bibfnamefont {T.}~\bibnamefont
  {Nakajima}}, \bibinfo {author} {\bibfnamefont {S.}~\bibnamefont {Gao}},
  \bibinfo {author} {\bibfnamefont {L.}~\bibnamefont {Peng}}, \bibinfo {author}
  {\bibfnamefont {A.}~\bibnamefont {Kikkawa}}, \bibinfo {author} {\bibfnamefont
  {T.}~\bibnamefont {Kurumaji}}, \bibinfo {author} {\bibfnamefont
  {M.}~\bibnamefont {Kriener}}, \bibinfo {author} {\bibfnamefont
  {Y.}~\bibnamefont {Yamasaki}}, \bibinfo {author} {\bibfnamefont
  {H.}~\bibnamefont {Sagayama}}, \bibinfo {author} {\bibfnamefont
  {H.}~\bibnamefont {Nakao}}, \bibinfo {author} {\bibfnamefont
  {K.}~\bibnamefont {Ohishi}}, \bibinfo {author} {\bibfnamefont
  {K.}~\bibnamefont {Kakurai}}, \bibinfo {author} {\bibfnamefont
  {Y.}~\bibnamefont {Taguchi}}, \bibinfo {author} {\bibfnamefont
  {X.}~\bibnamefont {Yu}}, \bibinfo {author} {\bibfnamefont {T.}~\bibnamefont
  {Arima}},\ and\ \bibinfo {author} {\bibfnamefont {Y.}~\bibnamefont
  {Tokura}},\ }\bibfield  {title} {\bibinfo {title} {Skyrmion phase and
  competing magnetic orders on a breathing kagom\'{e} lattice},\ }\href
  {https://doi.org/10.1038/s41467-019-13675-4} {\bibfield  {journal} {\bibinfo
  {journal} {Nat. Commun.}\ }\textbf {\bibinfo {volume} {10}},\ \bibinfo
  {pages} {5831} (\bibinfo {year} {2019})}\BibitemShut {NoStop}%
\bibitem [{\citenamefont {Li}\ \emph {et~al.}(2019)\citenamefont {Li},
  \citenamefont {Ding}, \citenamefont {Chen}, \citenamefont {Li}, \citenamefont
  {Hou}, \citenamefont {Liu}, \citenamefont {Zhang}, \citenamefont {Xi},
  \citenamefont {Wu},\ and\ \citenamefont {Wang}}]{li_large_2019}%
  \BibitemOpen
  \bibfield  {author} {\bibinfo {author} {\bibfnamefont {H.}~\bibnamefont
  {Li}}, \bibinfo {author} {\bibfnamefont {B.}~\bibnamefont {Ding}}, \bibinfo
  {author} {\bibfnamefont {J.}~\bibnamefont {Chen}}, \bibinfo {author}
  {\bibfnamefont {Z.}~\bibnamefont {Li}}, \bibinfo {author} {\bibfnamefont
  {Z.}~\bibnamefont {Hou}}, \bibinfo {author} {\bibfnamefont {E.}~\bibnamefont
  {Liu}}, \bibinfo {author} {\bibfnamefont {H.}~\bibnamefont {Zhang}}, \bibinfo
  {author} {\bibfnamefont {X.}~\bibnamefont {Xi}}, \bibinfo {author}
  {\bibfnamefont {G.}~\bibnamefont {Wu}},\ and\ \bibinfo {author}
  {\bibfnamefont {W.}~\bibnamefont {Wang}},\ }\bibfield  {title} {\bibinfo
  {title} {Large topological {Hall} effect in a geometrically frustrated kagome
  magnet {Fe$_3$Sn$_2$}},\ }\href {https://doi.org/10.1063/1.5088173}
  {\bibfield  {journal} {\bibinfo  {journal} {Appl. Phys. Lett.}\ }\textbf
  {\bibinfo {volume} {114}},\ \bibinfo {pages} {192408} (\bibinfo {year}
  {2019})}\BibitemShut {NoStop}%
\bibitem [{\citenamefont {Khanh}\ \emph {et~al.}(2020)\citenamefont {Khanh},
  \citenamefont {Nakajima}, \citenamefont {Yu}, \citenamefont {Gao},
  \citenamefont {Shibata}, \citenamefont {Hirschberger}, \citenamefont
  {Yamasaki}, \citenamefont {Sagayama}, \citenamefont {Nakao}, \citenamefont
  {Peng}, \citenamefont {Nakajima}, \citenamefont {Takagi}, \citenamefont
  {Arima}, \citenamefont {Tokura},\ and\ \citenamefont {Seki}}]{Khanh2020}%
  \BibitemOpen
  \bibfield  {author} {\bibinfo {author} {\bibfnamefont {N.~D.}\ \bibnamefont
  {Khanh}}, \bibinfo {author} {\bibfnamefont {T.}~\bibnamefont {Nakajima}},
  \bibinfo {author} {\bibfnamefont {X.}~\bibnamefont {Yu}}, \bibinfo {author}
  {\bibfnamefont {S.}~\bibnamefont {Gao}}, \bibinfo {author} {\bibfnamefont
  {K.}~\bibnamefont {Shibata}}, \bibinfo {author} {\bibfnamefont
  {M.}~\bibnamefont {Hirschberger}}, \bibinfo {author} {\bibfnamefont
  {Y.}~\bibnamefont {Yamasaki}}, \bibinfo {author} {\bibfnamefont
  {H.}~\bibnamefont {Sagayama}}, \bibinfo {author} {\bibfnamefont
  {H.}~\bibnamefont {Nakao}}, \bibinfo {author} {\bibfnamefont
  {L.}~\bibnamefont {Peng}}, \bibinfo {author} {\bibfnamefont {K.}~\bibnamefont
  {Nakajima}}, \bibinfo {author} {\bibfnamefont {R.}~\bibnamefont {Takagi}},
  \bibinfo {author} {\bibfnamefont {T.}~\bibnamefont {Arima}}, \bibinfo
  {author} {\bibfnamefont {Y.}~\bibnamefont {Tokura}},\ and\ \bibinfo {author}
  {\bibfnamefont {S.}~\bibnamefont {Seki}},\ }\bibfield  {title} {\bibinfo
  {title} {Nanometric square skyrmion lattice in a centrosymmetric tetragonal
  magnet},\ }\href {https://doi.org/10.1038/s41565-020-0684-7} {\bibfield
  {journal} {\bibinfo  {journal} {Nat. Nanotechnol.}\ }\textbf {\bibinfo
  {volume} {15}},\ \bibinfo {pages} {444} (\bibinfo {year} {2020})}\BibitemShut
  {NoStop}%
\bibitem [{\citenamefont {Yu}\ \emph {et~al.}(2014)\citenamefont {Yu},
  \citenamefont {Tokunaga}, \citenamefont {Kaneko}, \citenamefont {Zhang},
  \citenamefont {Kimoto}, \citenamefont {Matsui}, \citenamefont {Taguchi},\
  and\ \citenamefont {Tokura}}]{yu_biskyrmion_2014}%
  \BibitemOpen
  \bibfield  {author} {\bibinfo {author} {\bibfnamefont {X.~Z.}\ \bibnamefont
  {Yu}}, \bibinfo {author} {\bibfnamefont {Y.}~\bibnamefont {Tokunaga}},
  \bibinfo {author} {\bibfnamefont {Y.}~\bibnamefont {Kaneko}}, \bibinfo
  {author} {\bibfnamefont {W.~Z.}\ \bibnamefont {Zhang}}, \bibinfo {author}
  {\bibfnamefont {K.}~\bibnamefont {Kimoto}}, \bibinfo {author} {\bibfnamefont
  {Y.}~\bibnamefont {Matsui}}, \bibinfo {author} {\bibfnamefont
  {Y.}~\bibnamefont {Taguchi}},\ and\ \bibinfo {author} {\bibfnamefont
  {Y.}~\bibnamefont {Tokura}},\ }\bibfield  {title} {\bibinfo {title}
  {Biskyrmion states and their current-driven motion in a layered manganite},\
  }\href {https://doi.org/10.1038/ncomms4198} {\bibfield  {journal} {\bibinfo
  {journal} {Nat. Commun.}\ }\textbf {\bibinfo {volume} {5}},\ \bibinfo {pages}
  {3198} (\bibinfo {year} {2014})}\BibitemShut {NoStop}%
\bibitem [{\citenamefont {Batista}\ \emph {et~al.}(2016)\citenamefont
  {Batista}, \citenamefont {Lin}, \citenamefont {Hayami},\ and\ \citenamefont
  {Kamiya}}]{Batista2016}%
  \BibitemOpen
  \bibfield  {author} {\bibinfo {author} {\bibfnamefont {C.~D.}\ \bibnamefont
  {Batista}}, \bibinfo {author} {\bibfnamefont {S.-Z.}\ \bibnamefont {Lin}},
  \bibinfo {author} {\bibfnamefont {S.}~\bibnamefont {Hayami}},\ and\ \bibinfo
  {author} {\bibfnamefont {Y.}~\bibnamefont {Kamiya}},\ }\bibfield  {title}
  {\bibinfo {title} {Frustration and chiral orderings in correlated electron
  systems},\ }\href {https://doi.org/10.1088/0034-4885/79/8/084504} {\bibfield
  {journal} {\bibinfo  {journal} {Rep. Prog. Phys.}\ }\textbf {\bibinfo
  {volume} {79}},\ \bibinfo {pages} {084504} (\bibinfo {year}
  {2016})}\BibitemShut {NoStop}%
\bibitem [{\citenamefont {Niki}\ \emph {et~al.}(2015)\citenamefont {Niki},
  \citenamefont {Nakamura}, \citenamefont {Higa}, \citenamefont {Kuroshima},
  \citenamefont {Toji}, \citenamefont {Yogi}, \citenamefont {Nakamura},
  \citenamefont {Hedo}, \citenamefont {Nakama}, \citenamefont {{\={O}}nuki},\
  and\ \citenamefont {Harima}}]{Niki2015}%
  \BibitemOpen
  \bibfield  {author} {\bibinfo {author} {\bibfnamefont {H.}~\bibnamefont
  {Niki}}, \bibinfo {author} {\bibfnamefont {S.}~\bibnamefont {Nakamura}},
  \bibinfo {author} {\bibfnamefont {N.}~\bibnamefont {Higa}}, \bibinfo {author}
  {\bibfnamefont {H.}~\bibnamefont {Kuroshima}}, \bibinfo {author}
  {\bibfnamefont {T.}~\bibnamefont {Toji}}, \bibinfo {author} {\bibfnamefont
  {M.}~\bibnamefont {Yogi}}, \bibinfo {author} {\bibfnamefont {A.}~\bibnamefont
  {Nakamura}}, \bibinfo {author} {\bibfnamefont {M.}~\bibnamefont {Hedo}},
  \bibinfo {author} {\bibfnamefont {T.}~\bibnamefont {Nakama}}, \bibinfo
  {author} {\bibfnamefont {Y.}~\bibnamefont {{\={O}}nuki}},\ and\ \bibinfo
  {author} {\bibfnamefont {H.}~\bibnamefont {Harima}},\ }\bibfield  {title}
  {\bibinfo {title} {Studies of $^{27}${Al} {NMR} in {EuAl}$_4$},\ }\href
  {https://doi.org/10.1088/1742-6596/592/1/012030} {\bibfield  {journal}
  {\bibinfo  {journal} {J. Phys.: Conf. Ser.}\ }\textbf {\bibinfo {volume}
  {592}},\ \bibinfo {pages} {012030} (\bibinfo {year} {2015})}\BibitemShut
  {NoStop}%
\bibitem [{THE()}]{THEnote}%
  \BibitemOpen
  \href@noop {} {}\bibinfo {note} {{In} the momentum-space scenario, the term
  `anomalous Hall effect' is often used in the literature to denote any
  contribution other than the ordinary Hall effect. Here, we use the term
  `topological Hall effect' to denote the contribution in addition to the
  ordinary- and the \emph{conventional} anomalous Hall effect, as determined
  from magnetization measurements. In other words, the topological Hall effect
  we discuss here is the topologically nontrivial part of the anomalous Hall
  effect.}\BibitemShut {Stop}%
\bibitem [{\citenamefont {Nakatsuji}\ \emph {et~al.}(2015)\citenamefont
  {Nakatsuji}, \citenamefont {Kiyohara},\ and\ \citenamefont
  {Higo}}]{nakatsuji_large_2015}%
  \BibitemOpen
  \bibfield  {author} {\bibinfo {author} {\bibfnamefont {S.}~\bibnamefont
  {Nakatsuji}}, \bibinfo {author} {\bibfnamefont {N.}~\bibnamefont
  {Kiyohara}},\ and\ \bibinfo {author} {\bibfnamefont {T.}~\bibnamefont
  {Higo}},\ }\bibfield  {title} {\bibinfo {title} {Large anomalous {Hall}
  effect in a non-collinear antiferromagnet at room temperature},\ }\href
  {https://doi.org/10.1038/nature15723} {\bibfield  {journal} {\bibinfo
  {journal} {Nature}\ }\textbf {\bibinfo {volume} {527}},\ \bibinfo {pages}
  {212} (\bibinfo {year} {2015})}\BibitemShut {NoStop}%
\bibitem [{\citenamefont {Ikhlas}\ \emph {et~al.}(2017)\citenamefont {Ikhlas},
  \citenamefont {Tomita}, \citenamefont {Koretsune}, \citenamefont {Suzuki},
  \citenamefont {Nishio-Hamane}, \citenamefont {Arita}, \citenamefont {Otani},\
  and\ \citenamefont {Nakatsuji}}]{ikhlas_large_2017}%
  \BibitemOpen
  \bibfield  {author} {\bibinfo {author} {\bibfnamefont {M.}~\bibnamefont
  {Ikhlas}}, \bibinfo {author} {\bibfnamefont {T.}~\bibnamefont {Tomita}},
  \bibinfo {author} {\bibfnamefont {T.}~\bibnamefont {Koretsune}}, \bibinfo
  {author} {\bibfnamefont {M.-T.}\ \bibnamefont {Suzuki}}, \bibinfo {author}
  {\bibfnamefont {D.}~\bibnamefont {Nishio-Hamane}}, \bibinfo {author}
  {\bibfnamefont {R.}~\bibnamefont {Arita}}, \bibinfo {author} {\bibfnamefont
  {Y.}~\bibnamefont {Otani}},\ and\ \bibinfo {author} {\bibfnamefont
  {S.}~\bibnamefont {Nakatsuji}},\ }\bibfield  {title} {\bibinfo {title} {Large
  anomalous {Nernst} effect at room temperature in a chiral antiferromagnet},\
  }\href {https://doi.org/10.1038/nphys4181} {\bibfield  {journal} {\bibinfo
  {journal} {Nat. Phys.}\ }\textbf {\bibinfo {volume} {13}},\ \bibinfo {pages}
  {1085} (\bibinfo {year} {2017})}\BibitemShut {NoStop}%
\bibitem [{\citenamefont {Suzuki}\ \emph {et~al.}(2016)\citenamefont {Suzuki},
  \citenamefont {Chisnell}, \citenamefont {Devarakonda}, \citenamefont {Liu},
  \citenamefont {Feng}, \citenamefont {Xiao}, \citenamefont {Lynn},\ and\
  \citenamefont {Checkelsky}}]{suzuki_large_2016}%
  \BibitemOpen
  \bibfield  {author} {\bibinfo {author} {\bibfnamefont {T.}~\bibnamefont
  {Suzuki}}, \bibinfo {author} {\bibfnamefont {R.}~\bibnamefont {Chisnell}},
  \bibinfo {author} {\bibfnamefont {A.}~\bibnamefont {Devarakonda}}, \bibinfo
  {author} {\bibfnamefont {Y.-T.}\ \bibnamefont {Liu}}, \bibinfo {author}
  {\bibfnamefont {W.}~\bibnamefont {Feng}}, \bibinfo {author} {\bibfnamefont
  {D.}~\bibnamefont {Xiao}}, \bibinfo {author} {\bibfnamefont {J.~W.}\
  \bibnamefont {Lynn}},\ and\ \bibinfo {author} {\bibfnamefont {J.~G.}\
  \bibnamefont {Checkelsky}},\ }\bibfield  {title} {\bibinfo {title} {Large
  anomalous {Hall} effect in a half-{Heusler} antiferromagnet},\ }\href
  {https://doi.org/10.1038/nphys3831} {\bibfield  {journal} {\bibinfo
  {journal} {Nat. Phys.}\ }\textbf {\bibinfo {volume} {12}},\ \bibinfo {pages}
  {1119} (\bibinfo {year} {2016})}\BibitemShut {NoStop}%
\bibitem [{\citenamefont {Guo}\ \emph {et~al.}(2018)\citenamefont {Guo},
  \citenamefont {Wu}, \citenamefont {Wu}, \citenamefont {Smidman},
  \citenamefont {Cao}, \citenamefont {Bostwick}, \citenamefont {Jozwiak},
  \citenamefont {Rotenberg}, \citenamefont {Liu}, \citenamefont {Steglich},\
  and\ \citenamefont {Yuan}}]{guo_evidence_2018}%
  \BibitemOpen
  \bibfield  {author} {\bibinfo {author} {\bibfnamefont {C.~Y.}\ \bibnamefont
  {Guo}}, \bibinfo {author} {\bibfnamefont {F.}~\bibnamefont {Wu}}, \bibinfo
  {author} {\bibfnamefont {Z.~Z.}\ \bibnamefont {Wu}}, \bibinfo {author}
  {\bibfnamefont {M.}~\bibnamefont {Smidman}}, \bibinfo {author} {\bibfnamefont
  {C.}~\bibnamefont {Cao}}, \bibinfo {author} {\bibfnamefont {A.}~\bibnamefont
  {Bostwick}}, \bibinfo {author} {\bibfnamefont {C.}~\bibnamefont {Jozwiak}},
  \bibinfo {author} {\bibfnamefont {E.}~\bibnamefont {Rotenberg}}, \bibinfo
  {author} {\bibfnamefont {Y.}~\bibnamefont {Liu}}, \bibinfo {author}
  {\bibfnamefont {F.}~\bibnamefont {Steglich}},\ and\ \bibinfo {author}
  {\bibfnamefont {H.~Q.}\ \bibnamefont {Yuan}},\ }\bibfield  {title} {\bibinfo
  {title} {Evidence for {Weyl} fermions in a canonical heavy-fermion semimetal
  {YbPtBi}},\ }\href {https://doi.org/10.1038/s41467-018-06782-1} {\bibfield
  {journal} {\bibinfo  {journal} {Nat. Commun.}\ }\textbf {\bibinfo {volume}
  {9}},\ \bibinfo {pages} {4622} (\bibinfo {year} {2018})}\BibitemShut
  {NoStop}%
\bibitem [{\citenamefont {Nayak}\ \emph {et~al.}(2016)\citenamefont {Nayak},
  \citenamefont {Fischer}, \citenamefont {Sun}, \citenamefont {Yan},
  \citenamefont {Karel}, \citenamefont {Komarek}, \citenamefont {Shekhar},
  \citenamefont {Kumar}, \citenamefont {Schnelle}, \citenamefont {K{\"u}bler},
  \citenamefont {Felser},\ and\ \citenamefont {Parkin}}]{nayak_large_2016}%
  \BibitemOpen
  \bibfield  {author} {\bibinfo {author} {\bibfnamefont {A.~K.}\ \bibnamefont
  {Nayak}}, \bibinfo {author} {\bibfnamefont {J.~E.}\ \bibnamefont {Fischer}},
  \bibinfo {author} {\bibfnamefont {Y.}~\bibnamefont {Sun}}, \bibinfo {author}
  {\bibfnamefont {B.}~\bibnamefont {Yan}}, \bibinfo {author} {\bibfnamefont
  {J.}~\bibnamefont {Karel}}, \bibinfo {author} {\bibfnamefont {A.~C.}\
  \bibnamefont {Komarek}}, \bibinfo {author} {\bibfnamefont {C.}~\bibnamefont
  {Shekhar}}, \bibinfo {author} {\bibfnamefont {N.}~\bibnamefont {Kumar}},
  \bibinfo {author} {\bibfnamefont {W.}~\bibnamefont {Schnelle}}, \bibinfo
  {author} {\bibfnamefont {J.}~\bibnamefont {K{\"u}bler}}, \bibinfo {author}
  {\bibfnamefont {C.}~\bibnamefont {Felser}},\ and\ \bibinfo {author}
  {\bibfnamefont {S.~S.~P.}\ \bibnamefont {Parkin}},\ }\bibfield  {title}
  {\bibinfo {title} {Large anomalous {Hall} effect driven by a nonvanishing
  {Berry} curvature in the noncolinear antiferromagnet {Mn$_3$Ge}},\ }\href
  {https://doi.org/10.1126/sciadv.1501870} {\bibfield  {journal} {\bibinfo
  {journal} {Sci. Adv.}\ }\textbf {\bibinfo {volume} {2}},\ \bibinfo {pages}
  {e1501870} (\bibinfo {year} {2016})}\BibitemShut {NoStop}%
\end{thebibliography}%

\end{document}